\documentclass[reprint,twocolumn,english,superscriptaddress,aps,prb]{revtex4-2}
\usepackage[T1]{fontenc}
\usepackage[utf8]{inputenc}
\usepackage{babel}
\usepackage{bm}
\usepackage{amsmath}
\usepackage{graphicx}
\usepackage[unicode=true,pdfusetitle,
 bookmarks=true,bookmarksnumbered=false,bookmarksopen=false,
 breaklinks=false,pdfborder={0 0 1},backref=false,colorlinks=false]
 {hyperref}



\begin{document}

\title{Edge magnetism in transition metal dichalcogenide nanoribbons: Mean
field theory and determinant quantum Monte Carlo}

\author{Francisco M. O. Brito}
\affiliation{Department of Physics, University of York, YO10 5DD, York, United Kingdom}

\author{Linhu Li}
\affiliation{Guangdong Provincial Key Laboratory of Quantum Metrology and Sensing and School of Physics and Astronomy,
Sun Yat-Sen University (Zhuhai Campus), Zhuhai 519082, China}

\author{João M. V. P. Lopes}
\affiliation{Centro de Física das Universidades do Minho e Porto, Departamento
de Física e Astronomia, Faculdade de Ciências, Universidade do Porto,
4169-007 Porto, Portugal}

\author{Eduardo V. Castro}
\affiliation{Centro de Física das Universidades do Minho e Porto, Departamento
de Física e Astronomia, Faculdade de Ciências, Universidade do Porto,
4169-007 Porto, Portugal}
\affiliation{Beijing Computational Science Research Center, Beijing 100193, China}

\date{\today}
\begin{abstract}
Edge magnetism in zigzag transition metal dichalcogenide nanoribbons is studied using a three-band tight-binding model with local electron-electron interactions. Both mean field theory and the unbiased, numerically exact determinant quantum Monte Carlo method are applied. Depending on the edge filling, mean field theory predicts different phases: gapped spin dimer and antiferromagnetic phases appear for two specific  fillings, with a tendency towards metallic edge-ferromagnetism away from those fillings. Determinant quantum Monte Carlo simulations confirm the stability of the antiferromagnetic gapped phase at the same edge filling as mean field theory, despite being  sign-problematic for other fillings.  The obtained results  point to edge filling as yet another key ingredient to understand the observed magnetism in nanosheets. Moreover, the filling dependent edge magnetism gives rise to spin-polarized edge currents in zigzag nanoribbons which could be tuned through a back gate voltage, with possible applications to spintronics.
\end{abstract}
\maketitle

\section{introduction}

Transition metal dichalcogenides (TMDs) are prominent members of the
2D materials family \cite{manzeli_2d_2017} with numerous prospective
technological applications \cite{radisavljevic_single-layer_2011,koppens_photodetectors_2014,mak_photonics_2016}.
While monolayer graphene is gapless and its bilayer counterpart has
a tunable, but small gap of the order of a tenth of an eV \cite{zhang_direct_2009},
TMD monolayers are semiconducting, with intrinsic band gaps in excess
of $1$~eV \cite{mak_atomically_2010}. Since the direct band gap lies
in the visible frequency range, 
these semiconducting analogues of graphene are promising for optoelectronic applications \cite{tian_optoelectronic_2016,khan_optoelectronics_2018,SPE2020}.
TMDs are also promising in the rapidly growing fields of spin- and
valleytronics \cite{xiao_coupled_2012,schaibley_valleytronics_2016,enaldiev_edge_2017,ciccarino_dynamics_2018,zhou_spin-orbit_2019},
where it is particularly important to manipulate the electronic spin
and valley degrees of freedom \cite{PhysRevLett.123.096803}.

The presence of one-dimensional edges is a distinctive feature of any 2D material. The reduced dimensionality gives rise to unique properties which are not present in the bulk. 
Zigzag graphene nanoribbons (zGNRs) are known examples where low energy edge states appear. 
In the tight-binding picture,
these correspond to bands close to the Fermi energy that become flatter
and flatter as the width of the ribbon is increased. In Ref.~\cite{fujita_peculiar_1996},
mean field theory (MFT) was used to study electron-electron interactions
in zGNRs, revealing the possibility of spontaneous magnetic order
at the edges. Subsequent studies unraveled the rich
physics of these edge-states, supporting the existence of the magnetic
phase predicted using MFT, and unveiling electronic properties such
as half-metallicity \cite{wakabayashi_spin_1998,yamashiro_spin-_2003,son_half-metallic_2006,son_energy_2006,rudberg_nonlocal_2007,hod_enhanced_2007,fernandez-rossier_prediction_2008}. 
Despite the successful fabrication of graphene nanoribbons \cite{Jia2009,JZW+09}, the observation of magnetized zigzag edges is limited to the detection of spin-split edge bands using scanning tunneling microscopy \cite{crommieNatPhys2011}. Long-range magnetic order remains elusive \cite{NST+12}, and zGNR fabrication alternatives \cite{Chen2021} as well as new strategies to enhance edge magnetism are currently being explored \cite{Kaxiras2021}.

Similarly to graphene, TMDs can also be synthesized in the form of nanoribbons, as recently demonstrated through a variety of methods \cite{chen_fabrication_2017,Cheng2017,Li2018,Wang2019,Yang2019,Munkhbat2020}.
However, contrary to graphene, there is ample experimental evidence of edge-magnetic ordering
on few-layer TMD nanostructures \cite{ISI:000326898500001,gao2013ferromagnetism,ISI:000347245500034,ISI:000344531100001,ISI:000337140800041,ISI:000418492500019,ISI:000457723300001,ISI:000405591300006}. In ultrathin $\text{Mo}\text{S}_2$ and $\text{W}\text{S}_2$ nanosheets, ferromagnetic order sets in even at room temperature \cite{ISI:000326898500001,gao2013ferromagnetism,ISI:000347245500034,ISI:000344531100001,ISI:000337140800041}. The onset of magnetic order has been attributed to the presence of zigzag edges and/or structural defects such as grain boundaries or vacancies related to the synthesis process.
The ferromagnetic behavior found in few-layer $\text{Mo}\text{S}_2$ nanomeshes  \cite{ISI:000418492500019} supports the idea that the zigzag edge contribution is dominant: on one hand, the dependence on interpore distance mimics the dependence on the zigzag nanoribbon width; on the other,  ferromagnetism is absent in samples without annealing, where the proportion of as-grown defects compared to zigzag edges is higher.
Contrasting with exfoliated nanosheets, for which clear signs of ferromagnetism are observed, 
pristine TMDs, such as $\text{Mo}\text{S}_2$ in its three-dimensional form, are diamagnetic \cite{ISI:000418492500019}. Moreover, mono-/bi-layer $\text{Mo}\text{S}_2$ nanosheets show enhanced room temperature ferromagnetism attributed to an increased density of zigzag edges and/or defects \cite{ISI:000405591300006}, while bulk monolayers are only spin-valley polarized upon doping \cite{braz_valley-polarized_2018}.

On the theory side, extensive work based on density functional theory (DFT) calculations have predicted both metallic behavior and ferromagnetism at the edges of zigzag TMD nanoribbons (zTMDNRs) \cite{bollinger_one-dimensional_2001,bollinger_atomic_2003,chen08,vojvodic_magnetic_2009,botello-mendez_metallic_2009,magDFT2012,kou_tuning_2012,Lopez-Urias2014,Cui2017,Vancso2019}. 
These calculations indicate that the energy difference between ferromagnetic and antiferromagnetic spin ordering at the edges is around tens of meV \cite{Lopez-Urias2014,Vancso2019}. Such a small energy difference casts some doubt on what is the thermodynamically stable phase and also indicates that magnetic ordering in zTMDNRs may be sensitive to external perturbations, such as a back gate voltage, which in turn changes the edge filling. 
A realistic tight-binding parametrization with a mean field decoupling of the Hubbard interaction was recently used to test the stability of edge magnetism against disorder in zigzag $\text{Mo}\text{S}_2$ nanoribbons \cite{Vancso2019}. However, the sensitivity to the filling of the edge was not considered.
In zGNRs and also phosphorene nanoribbons, edge magnetism has been studied using the unbiased, numerically exact determinant Quantum
Monte Carlo (DetQMC) method \cite{yang_room-temperature_2016,feldner_magnetism_2010,feldner_dynamical_2011,raczkowski_interplay_2017,yang_strain-tuning_2017}. 
The results for zGNRs further support the emergence of edge-magnetic order from
electron-electron interactions and it is possible to make a direct
comparison between DetQMC and MFT results. 
The DetQMC method overcomes the limitations of the approximate local or semi-local functionals used in DFT calculations \cite{Shin2021}, and also the typical overestimation of long-range order in MFT \cite{feldner_magnetism_2010}. To the best of our knowledge, unbiased, numerically exact approaches have not yet been applied to study the magnetism of zTMDNRs. 

In this work, we study the magnetism of zTMDNRs using MFT and DetQMC \cite{hirsch_discrete_1983,blankenbecler_monte_1981,hanke_electronic_1993}, based on a widely used three-band tight-binding model \cite{liu_three-band_2013,chu_spin-orbit-coupled_2014}, to which we add electron-electron interactions. Within a minimal intraorbital Hubbard model, we find evidence for the existence of magnetic order at the zigzag edges from both MFT and DetQMC, analogously to zGNRs. 
MFT provides evidence for the existence of a metallic ferromagnetic phase and two gapped phases with antiferromagnetic order, depending on the edge filling.
This result is confirmed within MFT using a more elaborate model --- suitable for transition metal atoms --- which considers multiorbital interactions. The DetQMC results corroborate the existence of one of the gapped phases predicted with MFT for a specific edge filling.
For other edge fillings, the DetQMC algorithm suffers from the sign problem and a direct comparison with MFT is not possible. 

The remainder of this paper is organized as follows. In Sec.~\ref{sec:minimal-3-band-hubbard},
we present our minimal model. We also give a brief overview of the MFT and DetQMC methods in the context of the used minimal model. In Sec.~\ref{sec:results}, we present our
MFT and DetQMC results separately, closing the section with a critical comparison of the two methods. Conclusions are presented in Sec.~\ref{sec:conclusions}. 

\section{model and methods\label{sec:minimal-3-band-hubbard}}

\begin{figure}[ht]
\hspace{2mm}\includegraphics[scale=0.08]{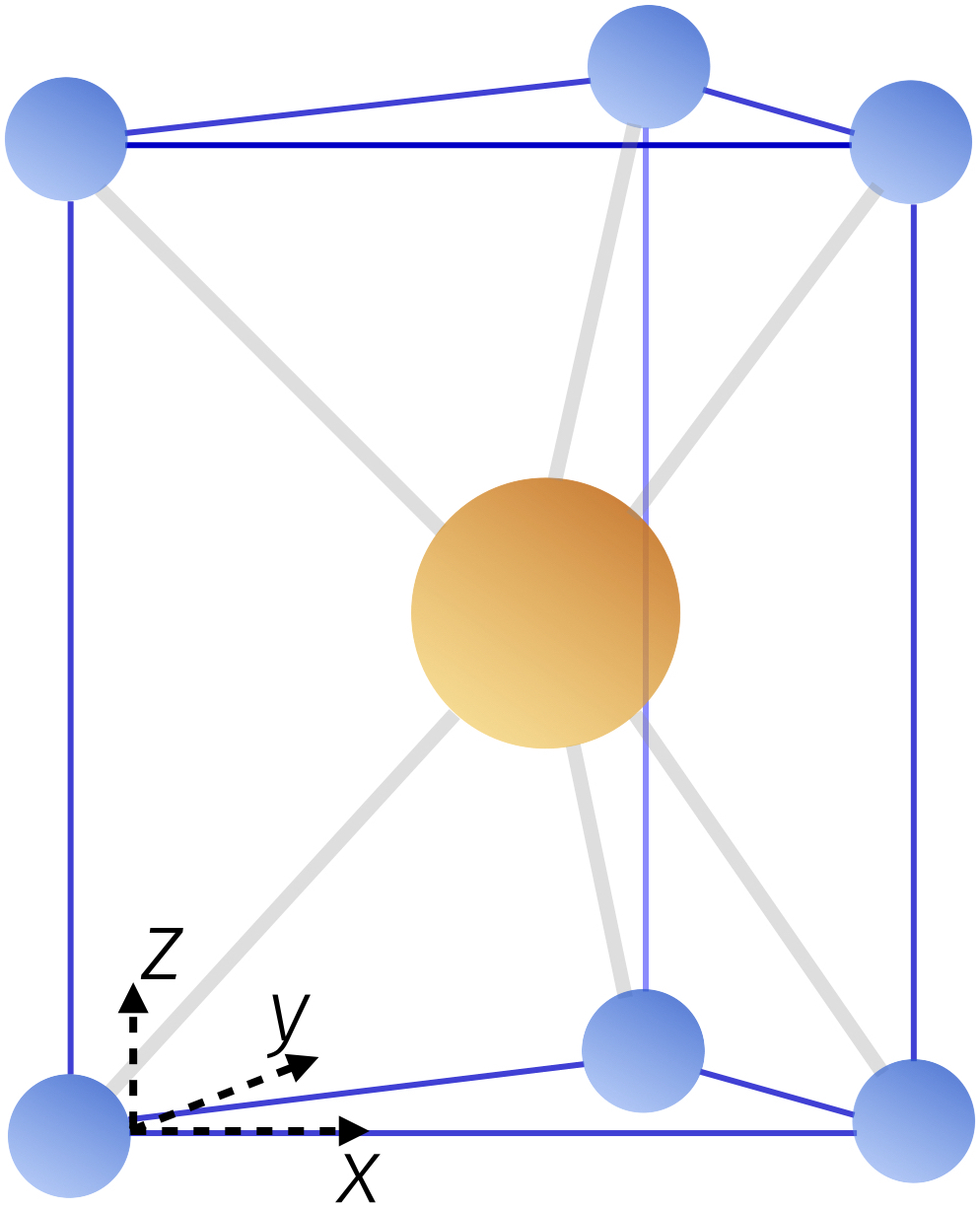}\hspace{5mm}\includegraphics[scale=0.16]{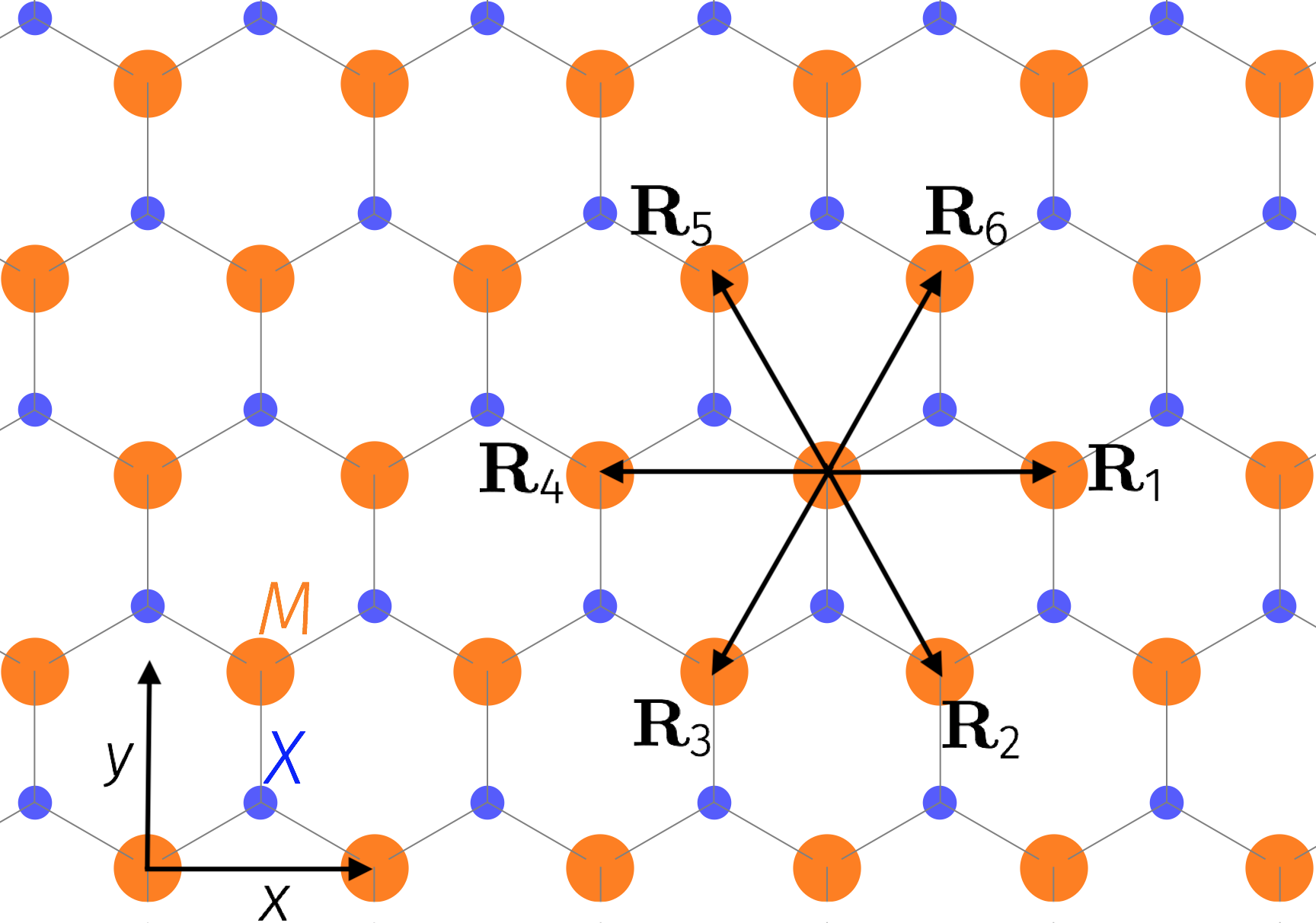}\caption{(Left) Schematic for the structure coordination in the
trigonal prismatic (2H) phase of a TMD. (Right) Projection of the 2H structure onto the $xy$ plane, yielding a honeycomb lattice. Hopping terms directly involving the X atoms are neglected in our model. Thus we identify only the relevant nearest $M$ neighbors by the vectors $\bm{R}_{i=1,2,...,6}$.  This lattice represents part of the nanoribbon with a width of 5 $M$ atoms we shall consider later. Each row of the ribbon is defined as a set of $M$ atoms for which $y$ is constant. 
\label{fig:NNneighborsTMDmonolayer}}
\end{figure}

Group 6 TMDs contain transition metals $\text{M}=\text{Mo},\text{W}$,
and chalcogens $\text{X}=\text{S},\text{Se},\text{Te}$, in a 1:2 proportion, and thus have the chemical formula $\text{M}\text{X}_{2}$.
In the monolayer form, $M$ atoms are arranged in a triangular lattice, sandwiched between two layers of $\text{X}$ atoms. 
The most common stacking structure --- shown in the left panel of Fig.~\ref{fig:NNneighborsTMDmonolayer} 
--- is denoted trigonal prismatic (2H). In this work, we consider the planar honeycomb lattice corresponding to the 2H unit cell depicted in the top-down view of the right panel of Fig.~\ref{fig:NNneighborsTMDmonolayer}.

We consider a minimal intraorbital Hubbard model based on the three-band tight-binding
model of Ref.~\cite{liu_three-band_2013}:
\begin{equation}
\mathcal{H}=\sum_{\substack{\left\langle i,j \right\rangle,\sigma,\\
\alpha,\beta }}
c_{i,\alpha,\sigma}^{\dagger}K_{\alpha\beta}(\bm{R}_{i j})c_{j,\beta,\sigma}+U\sum_{i,\alpha}n_{i,\alpha,\uparrow}n_{i,\alpha,\downarrow},\label{eq:hubbard}
\end{equation}
where $\left\langle i,j\right\rangle$ are nearest-neighbor sites on the triangular ($M$ atom) lattice, $c_{i, \alpha, \sigma}^{\dagger}$, $c_{j, \beta, \sigma}^{\dagger}$ are electron creation operators on lattice sites $i,j$,   $M$ atom orbitals $\alpha,\beta=d_{z^{2}}, d_{x y}, d_{x^{2}-y^{2}}$ and spin $\sigma=\uparrow, \downarrow$, $n=c^{\dagger}c$ is the number operator and $U$ is the Hubbard interaction.
We use the hopping parameters $K_{\alpha\beta}(\bm{R}_{i j})$ obtained with the generalized-gradient
approximation in Ref.~\cite{liu_three-band_2013}.
To mimic the geometry of the nanoribbon, we consider periodic boundary conditions (PBCs) along the longitudinal ($x$) direction and open boundary conditions (OBCs) along the transverse ($y$) direction (see right panel of Fig.~\ref{fig:NNneighborsTMDmonolayer}).
In order to capture potential multiorbital effects, one needs to go beyond our minimal Hubbard model by adding the following terms to the Hamiltonian of Eq.~(\ref{eq:hubbard}): an  interorbital on-site interaction term ($U^\prime{}$), a Hund term ($J$) and a pair-hopping term ($J^\prime{}$),
\begin{align}\label{eq:add_terms0}
\mathcal{H}_\text{inter-orb.} = \frac{U^\prime{}}{2} &\sum_{\substack{i,\alpha\neq\beta \\ \sigma, \sigma^\prime{}}}   n_{i,\alpha,\sigma} n_{i,\beta,\sigma^\prime{}}, \\
\label{eq:add_terms1}
\mathcal{H}_\text{Hund} =\frac{J}{2} &\sum_{\substack{i,\alpha\neq\beta \\ \sigma, \sigma^\prime{}}} c_{i,\alpha,\sigma}^\dagger c_{i,\beta,\sigma^\prime{}}^\dagger c_{i,\alpha,\sigma^\prime{}} c_{i,\beta,\sigma},  \\
\label{eq:add_terms2}
\mathcal{H}_\text{pair hopp.} =\frac{J^\prime{}}{2} &\sum_{\substack{i,\alpha\neq\beta \\ \sigma \neq \sigma^\prime{}}} c_{i,\alpha,\sigma}^\dagger c_{i,\alpha,\sigma^\prime{}}^\dagger c_{i,\beta,\sigma^\prime{}} c_{i,\beta,\sigma}. 
\end{align}
Assuming rotational invariance, out of the four parameters characterizing the
on-site interaction, only two are independent. For $d$-orbitals, the following relation holds:  
$J^\prime{} = J = (U-U^\prime{})/2$ \cite{DAGOTTO20011}.
Despite the lack of full rotational symmetry in TMDs, deviations from this relation are not severe \cite{Held-PhysRevB.90.165105-2014}, and we still use it here.

\begin{figure}[ht]
\includegraphics[width=8.5cm]{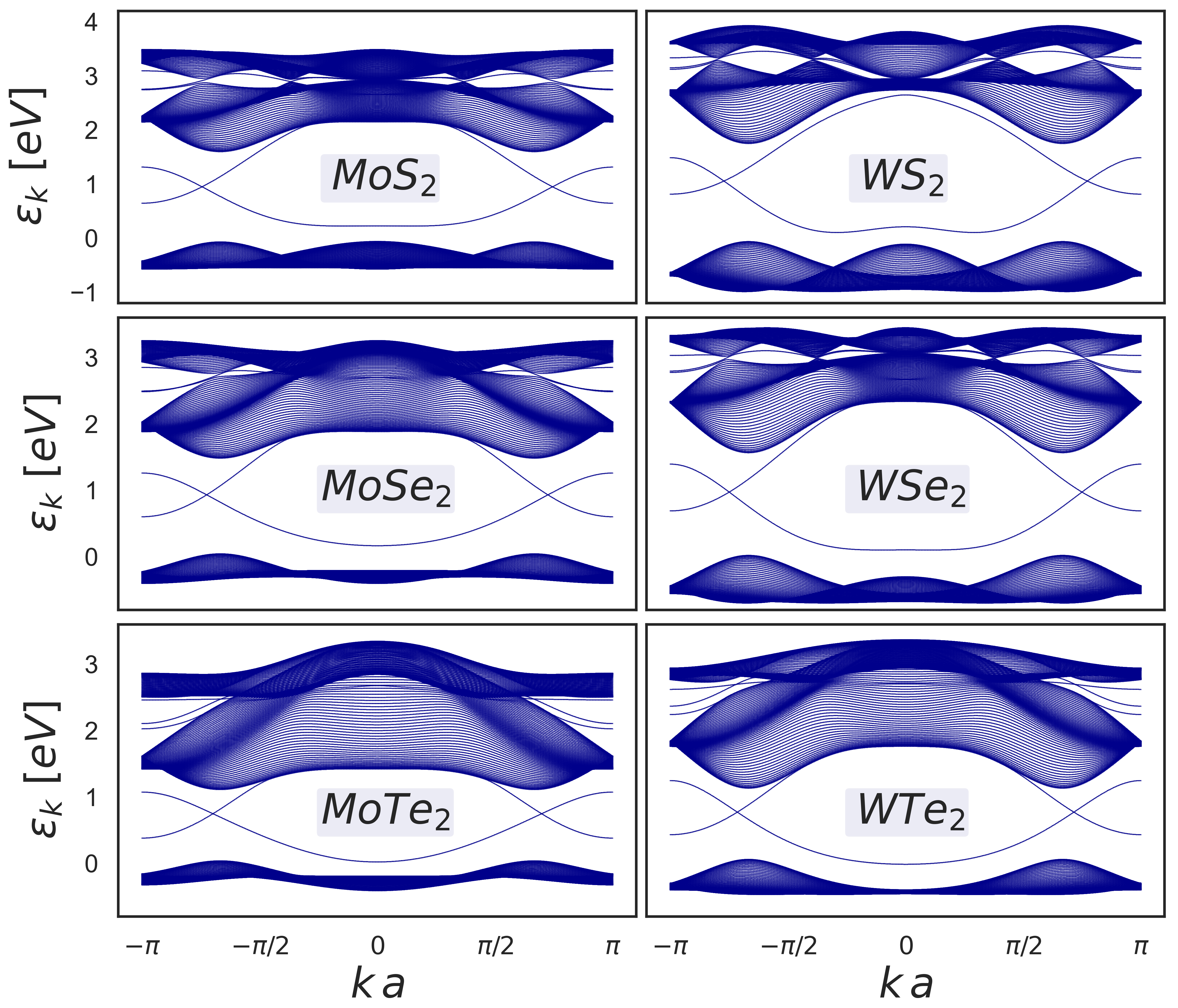}
\caption{Band structures of the three-band tight-binding model, i.e. Eq.~\eqref{eq:hubbard} with $U=0$, for various infinitely long zTMDNRs with a width of 64 $M$ atoms.\label{fig:Band-structures-TMDNR}}
\end{figure}

In Fig.~\ref{fig:Band-structures-TMDNR}, we show the energy bands obtained with the noninteracting three-band tight-binding model for zTMDNRs with a width of 64~$M$ atoms. The
two in-gap, spin-degenerate bands at around $1$~eV correspond to states localized at the $M$ and $X$ terminated edges. It should be noted that the minimal model considered here correctly 
reproduces the edge bands derived from $d_{z^{2}}$, $d_{x y}$, and $d_{x^{2}-y^{2}}$ orbitals
 \cite{liu_three-band_2013,chu_spin-orbit-coupled_2014}.
 
TMDs have a sizable spin-orbit coupling (SOC) ranging from several tens to a few hundreds of meV~\cite{liu_three-band_2013}.
Yet, as we shall see later, the minimum Hubbard interaction required for magnetic ordering is approximately an order of magnitude larger than the SOC energy scale. Thus, while SOC is a crucial feature of the real material, it can be safely neglected for our purposes. This relatively weak SOC (compared to $U$ required for magnetic ordering) further justifies the use of the rotationally invariant Coulomb interaction vertex of Eqs.(\ref{eq:add_terms0}-\ref{eq:add_terms2}). Another important remark is that $SU(2)$ spin-rotation symmetry is broken by
SOC and the Mermin-Wagner theorem does not apply~\cite{mermin_absence_1966}.
Therefore edge-magnetic ordering at finite temperature is not ruled
out in TMDNRs.

\subsection{Mean field theory\label{subsec:mean field-theory}}

The Gibbs-Bogolyubov-Feynman inequality states that
the variational grand potential $\Omega_{V}$ --- computed with
a quadratic mean field Hamiltonian $\mathcal{H}_\text{MF}$ --- is an upper bound on the grand
potential $\Omega$ computed with a corresponding interacting Hamiltonian $\mathcal{H}$:
\begin{equation}
\Omega\le
\Omega_\text{MF}+\left\langle \mathcal{{H}}-\mathcal{{H}}_\text{MF}\right\rangle _\text{MF}
\equiv\Omega_{V},
\label{eq:variational_principle}
\end{equation}
where $\Omega_{M F}=-k_{B} T \log \operatorname{Tr}\left\{e^{-\beta\left(\mathcal{H}_{M F}-\mu \mathcal{N}_{e}\right)}\right\}$
is the mean field grand potential, $T$ is the temperature and $\beta = (k_B T)^{-1}$, $\mathcal{N}_{e}$ is the 
total electron number operator 
and $\left\langle \dots \right\rangle$ is a thermodynamical average with respect to $\mathcal{H}_\text{MF}$.
The chemical potential $\mu$ is set by fixing the electron density.

Starting from the mean field Hamiltonian family 
\begin{equation}
\mathcal{H}_{\text{MF}}=\mathcal{{H}}_\text{TB}+U\sum_{i,\alpha,\sigma}n_{i,\alpha,\sigma}f_{i,\alpha,-\sigma},\label{eq:mf_hubbard}
\end{equation}
where $\mathcal{{H}}_\text{TB}$ is the three-band tight-binding Hamiltonian and $f_{i,\alpha,\sigma}(\bm{\langle n\rangle})$ is a variational field, 
we apply the variational principle of Eq.~(\ref{eq:variational_principle})
to find the $\mathcal{H}_\text{MF}$ that best approximates the model Hamiltonian of  Eq.~(\ref{eq:hubbard}).
The variational field $\bm f$ that minimizes $\Omega_{V}$ is obtained by self-consistently solving the set of mean field equations
$f_{i,\alpha,\sigma}(\bm{\langle n\rangle})=\left\langle n_{i,\alpha,\sigma}\right\rangle \forall i,\alpha,\sigma $. 

\begin{figure}[ht]
\centering\includegraphics[width=8.7cm]{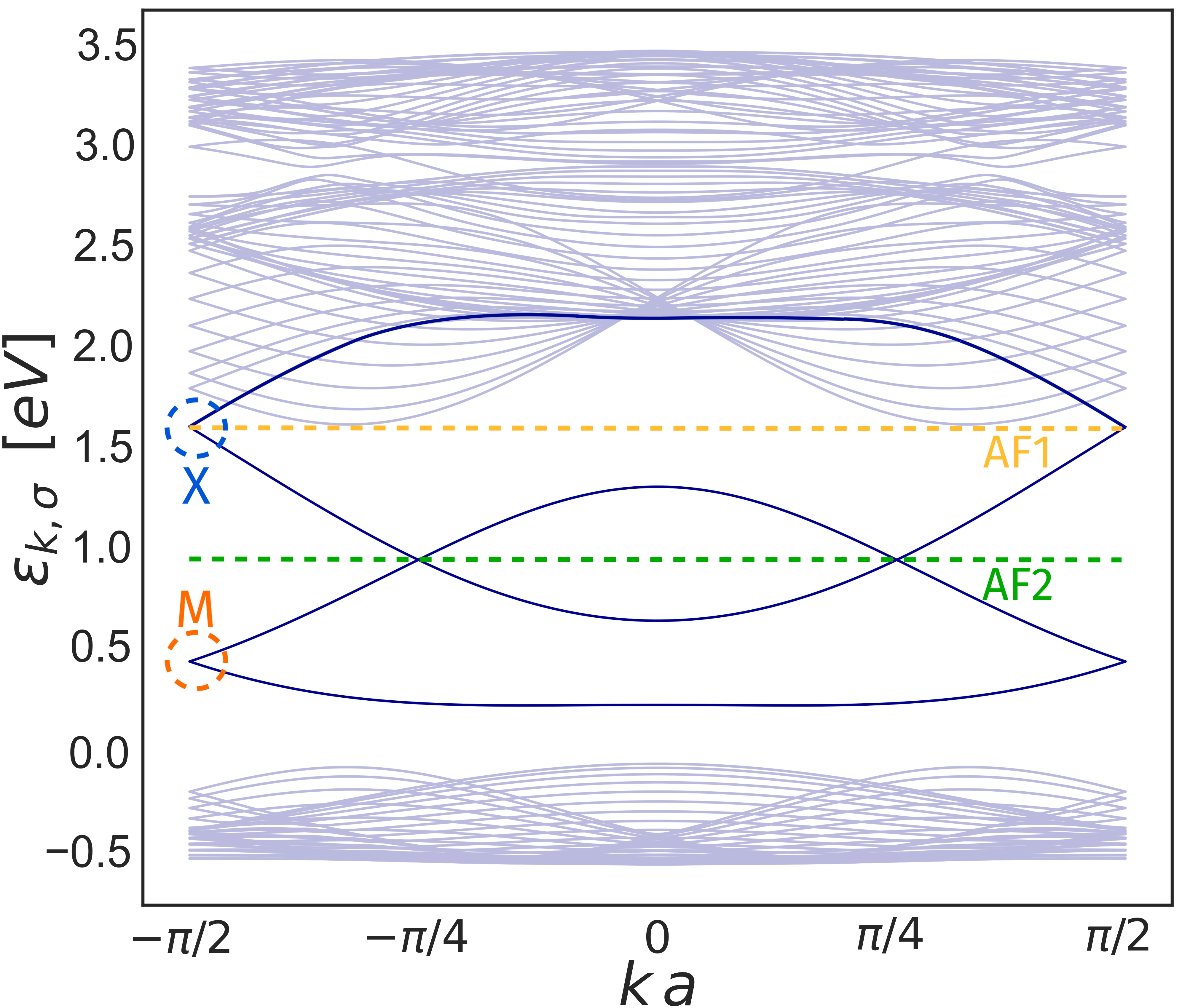}
\caption{Folded band structure of the three-band tight-binding model, i.e. Eq.~\eqref{eq:hubbard} with $U=0$, for a $\text{{Mo}}\text{{S}}_{2}$ nanoribbon
with a width of 20 $M$ atoms and a doubled unit cell in the longitudinal direction.
The bands corresponding to bulk states are faded. The dashed 
horizontal lines indicate band fillings where two types of antiferromagnetic order develop. \label{fig:free-bands}}
\end{figure}

To simplify the mean field analysis, we  assume that the mean field
 Hamiltonian of Eq.~(\ref{eq:mf_hubbard}) has translational symmetry along the longitudinal direction. 
 A discrete Fourier transform decouples the mean field Hamiltonian  into a set of effective one dimensional chains  in the transverse direction, one for each longitudinal quasimomentum $k$. 
 The choice of periodicity in the longitudinal direction restricts the self-consistent fields, and consequently the types of magnetic ordering that can be obtained by minimizing $\Omega_V$. 
 Let  $p$ be the number of $M$ atom columns in the unit cell. Taking $p=1$ only allows paramagnetic or ferromagnetic solutions, while 
$p=2$ also allows alternating spins (i.e. ferrimagnetic or antiferromagnetic
solutions). In order to capture more complex forms of magnetic ordering one has to consider larger periods. In particular, to detect the spin dimers we shall encounter later, one has to set $p=4$. 
Upon taking a unit cell with $p > 1$, the first Brillouin zone is reduced by a factor of $p$ and energy bands fold $p$ times. In Fig.~\ref{fig:free-bands}, we highlight the bands of the three-band tight-binding model with $p=2$. The bands labeled $M$, $X$ correspond to the $M$ and $X$ terminated
edges, respectively. The green and yellow lines indicate band fillings where --- as we shall see later --- gaps will be opened due to the Hubbard term in MFT. These gaps correspond to edge-dimer (AF2) and edge-antiferromagnetic phases (AF1).

In order to study the effect of the multiorbital terms of Eqs.~\eqref{eq:add_terms0}-\eqref{eq:add_terms2}, we further simplify our MFT approach by using a minimal set of variational fields. This set is obtained by making educated assumptions following the results obtained with the thorough MFT analysis of the intraorbital Hamiltonian of Eq.~(\ref{eq:mf_hubbard}). Firstly, we assume that the mean field solution is such that the bulk magnetization $\langle \hat{S}^z_{\,\,i,\alpha} \rangle \equiv \left\langle  n_{i,\alpha,\uparrow} - n_{i,\alpha,\downarrow} \right\rangle$ (for $i,\alpha$ on every row other than the two edge rows) vanishes. Then, we assume that magnetic order is primarily sensitive to electron spin. Thus, we ignore any orbital dependence. Finally, we compare the value of $\Omega_V$ in Eq.(\ref{eq:variational_principle}) for the following phases: paramagnetic, (anti)ferromagnetic on each of the edges --- labeled as Ferro-X, Ferro-M, AF-X, AF-M --- or on both --- labeled as Ferro-MX, AF-MX.

\subsection{Determinant Quantum Monte Carlo \label{subsec:DetQMC}}

DetQMC is commonly used to simulate interacting models of 2D nanostructures \cite{varney_quantum_2009,feldner_magnetism_2010, golor_quantum_2013,johnston_determinant_2013, cheng_strain-induced_2015, yang_room-temperature_2016, yang_strain-tuning_2017, raczkowski_interplay_2017}.
It is based on the Hubbard Stratonovich transformation, which allows one to map the Hubbard model onto a Hamiltonian of independent fermions coupled to a binary auxiliary field. 
Averages of quantum operators $\left\langle \mathcal{A} \right\rangle $ are evaluated by employing importance sampling over configurations
of the auxiliary field.  Each field configuration contributes to the expectation value with a weight that can be negative, leading to the fermion-sign problem. 
For sign-problematic models, the distribution of the sign variable is such that the variance of QMC estimators increases exponentially with the system size and the inverse temperature \cite{troyer_computational_2005, bai_numerical_2009}. 
In our case, a strong Hubbard interaction $U$ deems the model severely sign-problematic. The severity of the sign problem also varies with the chemical potential. The average of the sign distribution $\left\langle \text{sign} \right\rangle$ is a good measure of the severity of the sign problem for a given set of parameters $\beta,\mu, U, N$ (the latter being the total number of sites$\times$orbitals). As $\left\langle \text{sign} \right\rangle \rightarrow 0$, the sign problem becomes more severe and the QMC estimators are no longer reliable.

We measure the $\hat{S}^{z}$ spin-spin correlator between sites $i$ and $j$ with DetQMC: 
\begin{equation}
C(\bm{R}_{i},\bm{R}_{j}) =\sum_{\alpha,\beta}\left\langle (n_{i, \alpha,\uparrow}-n_{i, \alpha,\downarrow}) ( n_{j, \beta,\uparrow}-n_{j, \beta,\downarrow}) \right\rangle.
\end{equation}
Notice that we use the following definition throughout: $\hat{S}^{z} = n_\uparrow - n_\downarrow$. Translational invariance and mirror symmetry are used  to maximize the amount of information extracted from the measured values of the observable $C(\bm{R}_{i},\bm{R}_{j})$.
The discrete Fourier transform of the spin-spin correlator,  known
as the magnetic structure factor,
\begin{equation}
S(\bm{q})=\frac{1}{N}\sum_{i,j}e^{i\bm{q}\cdot(\bm{R}_{j}-\bm{R}_{i})}C(\bm{R}_{i},\bm{R}_{j}),\label{eq:structure_factor}
\end{equation}
is used to
carry out finite-size scaling analysis and probe the system for long-range order in the thermodynamic
limit. Peaks at $\bm q = (0, 0)$
and $\bm{q}=(\pi / a, \pi / a)$, where $a$ is the lattice constant, correspond to ferromagnetic and antiferromagnetic
order, respectively. Other types of order yield different peaks.

We inspect the TMD nanoribbon for edge magnetism by restricting
the sum in Eq.~(\ref{eq:structure_factor}) to the rows 
corresponding to the $\text{M}-$ and $\text{X}-$edges. If $N_x$ is the ribbon length, and setting $q_{x}=\pi$ to study  antiferromagnetic order, the structure factor for row $y$ can be written as:
\begin{equation}\label{eq:s_row}
S_{row}(\pi, y) = \frac{1}{N_x}\sum_{x_{i}, x_j=0}^{N_{x}-1}(-1)^{|x_{i}-x_{j}|}C(x_{i},y,x_{j},y).
\end{equation}

In practice, for
finite-size systems, we aim to obtain an estimate of the correlation
length $\xi$ and
compare it with $N_{x}$. When $\xi\ll N_{x}$, we are sufficiently
close to the thermodynamic limit to identify an ordered phase.
Due to translational invariance, we have that $C(x_{i},y,x_{j},y) = C(|x_{i} - x_{j}|, y)$. Defining $x \equiv | x_i - x_j | $, we hypothesize that
\begin{equation}
\frac{(-1)^{x}}{N_x} C(x, y)=f\bigg(\frac{x}{\xi}\bigg)+m_{s}^{2} (y),\label{eq:scaling_form}
\end{equation}
where $m_{s}^{\,2}(y)$ is the row-dependent staggered magnetization and $f$ is an integrable, monotonically decreasing function. Then, we may evaluate whether or not an ordered
phase appears by testing the consistency of our hypothesis. 

Replacing Eq.~(\ref{eq:scaling_form}) in Eq.~(\ref{eq:s_row}), we obtain
\begin{equation}
S_{row}(\pi,y)=\sum_{x=0}^{N_x-1}\bigg(f\Big(\frac{x}{\xi}\Big)+m_{s}^{\,2} (y) \bigg).
\end{equation}
If $\xi\ll N_{x}$, the quantity $(-1)^{x}C(x, y) / N_x$ becomes constant at long distances and converges to the squared staggered magnetization. Defining $A\equiv\sum_{x}f(x/\xi)$, we obtain
\begin{equation}
\frac{S_{row}(\pi,y)}{N_{x}}=\frac{A}{N_{x}}+m_{st}^{2}(y).\label{eq:extrapol}
\end{equation}

Equation (\ref{eq:extrapol}) can be used to obtain an estimate of the staggered
magnetizations in the thermodynamic limit.
We start by considering a temperature that
is low enough to find signs of magnetic ordering, but high enough to avoid convergence problems (see Fig.~\ref{fig:Spin-spin-correlators-U-beta}). Then,
we simulate systems with varying longitudinal length $N_{x}$, and
use Eq.~(\ref{eq:extrapol}) to extrapolate to the thermodynamic limit.

\section{results\label{sec:results}}

A central aspect of the present work is the study of magnetic instabilities in zTMDNRs as a function
of electron occupation of the edge bands. To be specific, we define the edge filling, $\nu_{edge}$, as the
fraction of electrons filling edge bands relative to the total number of available edge states
within the noninteracting three-band tight-binding model, i.e. $\nu_{edge} \in [0,1]$. 
Defining the spin-dependent electron density as $\left\langle n_\sigma \right\rangle =  N^{-1}\sum_{i, \alpha} \left\langle n_{i,\alpha, \sigma} \right\rangle $, where $N$ the total number of sites$\times$orbitals, such that $\left\langle n_\sigma \right\rangle \in [0,1]$, we may write the electron density as $\left\langle n \right\rangle = \left\langle n_\uparrow \right\rangle + \left\langle n_\downarrow \right\rangle$, which then ranges from 0 to 2. After simple algebra, the edge filling  can be written in terms of the electron density as
\begin{equation}\label{eq:nu_edge}
\nu_{edge} = \frac{3 N_y}{4} \left\langle n \right\rangle - \frac{1}{2} (N_y - 1),
\end{equation}
where $N_y$ is the width of the ribbon. For example, charge neutrality in this model corresponds
to $\left\langle n \right\rangle = 2/3$, which corresponds to $\nu_{edge}= 1/2$, i.e., half-filling
of the edge.

\subsection{Mean field theory: intraorbital interaction}

We start by presenting MFT results for the intraorbital Hubbard-$U$ model given by Eq.~\eqref{eq:hubbard}.
Only results for $\text{{Mo}\text{{S}}}_{2}$ parameters are shown, but we found  similar results for other TMDs of the family (see appendix \ref{sec:A}).

Long-range edge-magnetic order emerges as the intraorbital Hubbard interaction $U$ is increased. The type
of magnetic ordering depends on the edge filling $\nu_{edge}$. In particular, two gapped phases emerge: a spin-dimer phase (AF2) at $\nu_{edge}=1/2$ and an edge-antiferromagnetic phase (AF1) at $\nu_{edge} = 3/4$.
These two edge fillings are indicated by dashed lines in  Fig.~\ref{fig:free-bands}.
Both below $\nu_{edge} = 1/2$ and between the two gapped phases, MFT predicts edge-ferromagnetic order. 
As can be seen in Fig.~\ref{fig:free-bands}, for $\nu_{edge}>3/4$ bulk conduction bands start to be 
populated and the physics is no longer edge dominated.

\begin{figure}[ht]
\centering\includegraphics[width=8.3cm]{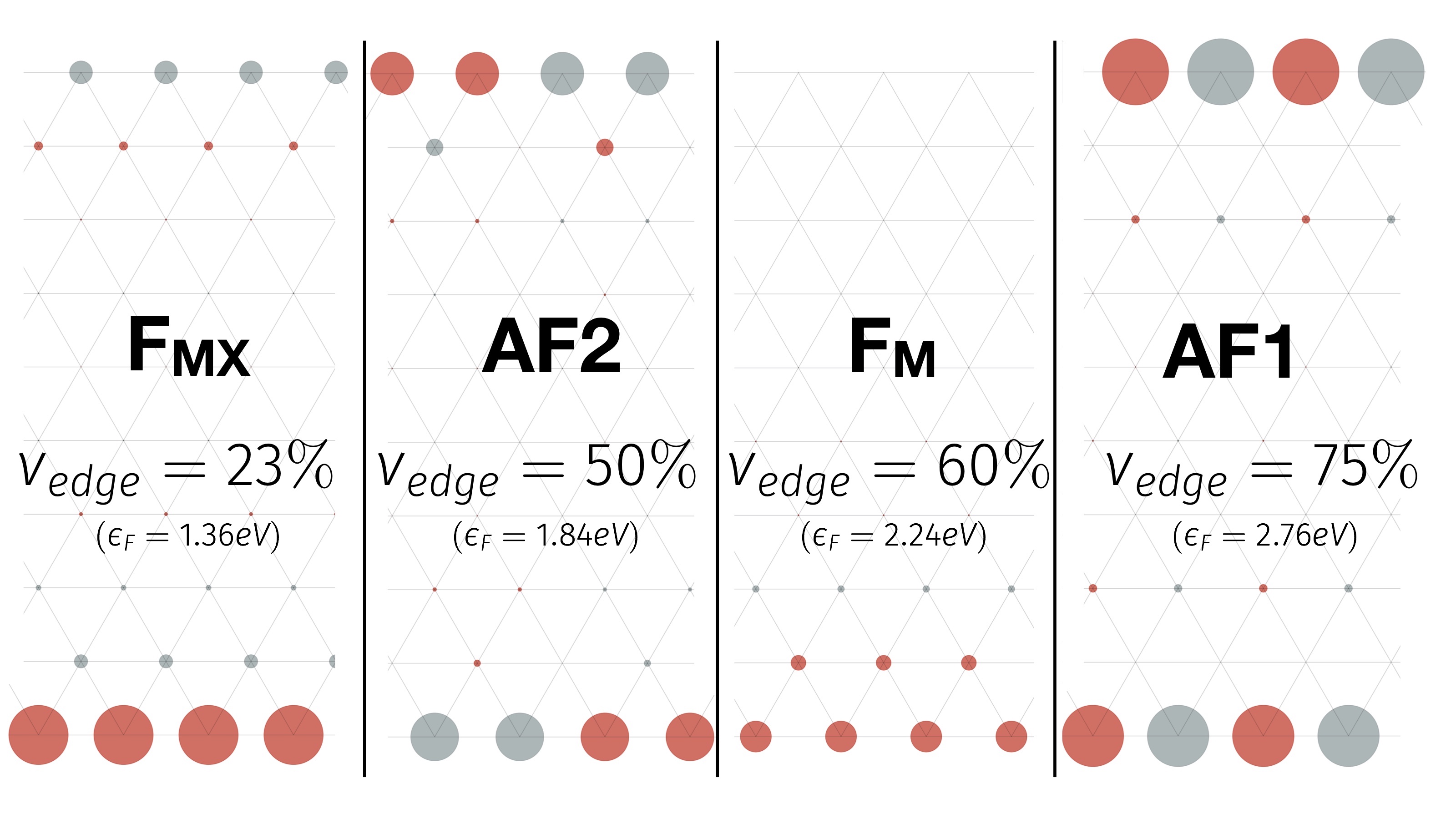}

\caption{Mean field magnetic ordering for $U=2.94\,\text{eV}$ at zero temperature for varying edge filling. 
The size of the circles indicates the magnitude of the local spin density. Red
corresponds to spin-up and grey to spin-down. Left to right: ferromagnetic
phase on both edges ($\text{{F}}_{\text{{MX}}}$); dimer phase (AF2);
ferromagnetic phase on the $\text{M}-$edge ($\text{{F}}_{\text{{M}}}$);
antiferromagnetic phase (AF1). \label{fig:mf_phases}}
\end{figure}

On the basis of the free edge bands shown in Fig.~\ref{fig:free-bands}, we can see that the gapped antiferromagnetic phases AF1 and AF2 are associated with the nesting vectors $Qa=\pi$ and $Qa=\pi/2$, respectively. Nesting favors gap opening instabilities such as these two types of antiferromagnetic ordering.
For other generic fillings, a splitting of the
spin-up and spin-down bands is preferred, which in turn induces Stoner-like edge-ferromagnetism. 
The three different types of edge-magnetic phases are shown in Fig.~\ref{fig:mf_phases},
where the profile of the local magnetization can be seen
along the rows of the ribbon. At the edges, it is higher in magnitude, decreasing rapidly and 
eventually vanishing in the bulk.

\begin{figure}[ht]
\centering
\includegraphics[width=8.5cm]{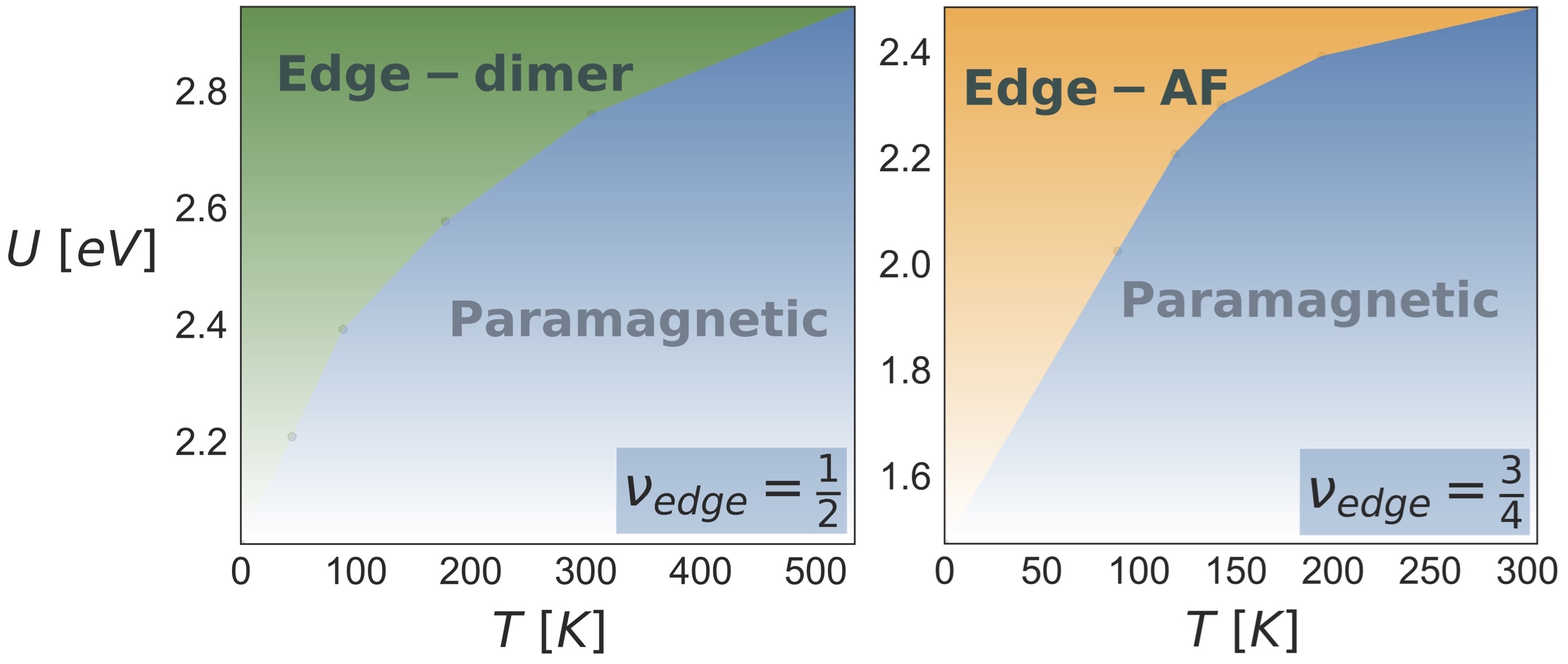}
\caption{$U-T$ mean field phase diagrams at: half edge filling (left); three-quarter
edge filling (right). As indicated, blueish regions correspond to the 
paramagnetic phase. A ribbon of width $N_y = 5$ $M$ atoms was used.\label{fig:mf_phase_diagram}}
\end{figure}

The U-T phase diagrams
for the antiferromagnetic (AF1) and dimer (AF2) phases are shown in Fig.~\ref{fig:mf_phase_diagram}.
It can be seen that these gapped phases exist at and below room temperature at the mean field level.
Although this stability might be overestimated in MFT, it suggests that DetQMC calculations are
worth doing for this model. Moreover, even though the Coulomb repulsion parameters are largely unknown for TMDs, the modest values of the Hubbard interaction obtained  are well within current parameter estimates \cite{Roldan2013}.
Let us also mention that, due to the asymmetry between the two edges, the critical $U$ for magnetic order to develop
is different for each of the two edges. In particular, for the AF1 phase only the $X$ edge is magnetized in the phase diagram
of Fig.~\ref{fig:mf_phase_diagram} (right panel). The $M$ edge becomes polarized only for $U\gtrsim 2.8$~eV, as will
be clear below.

\begin{figure}[ht]
\centering\includegraphics[width=8.2cm]{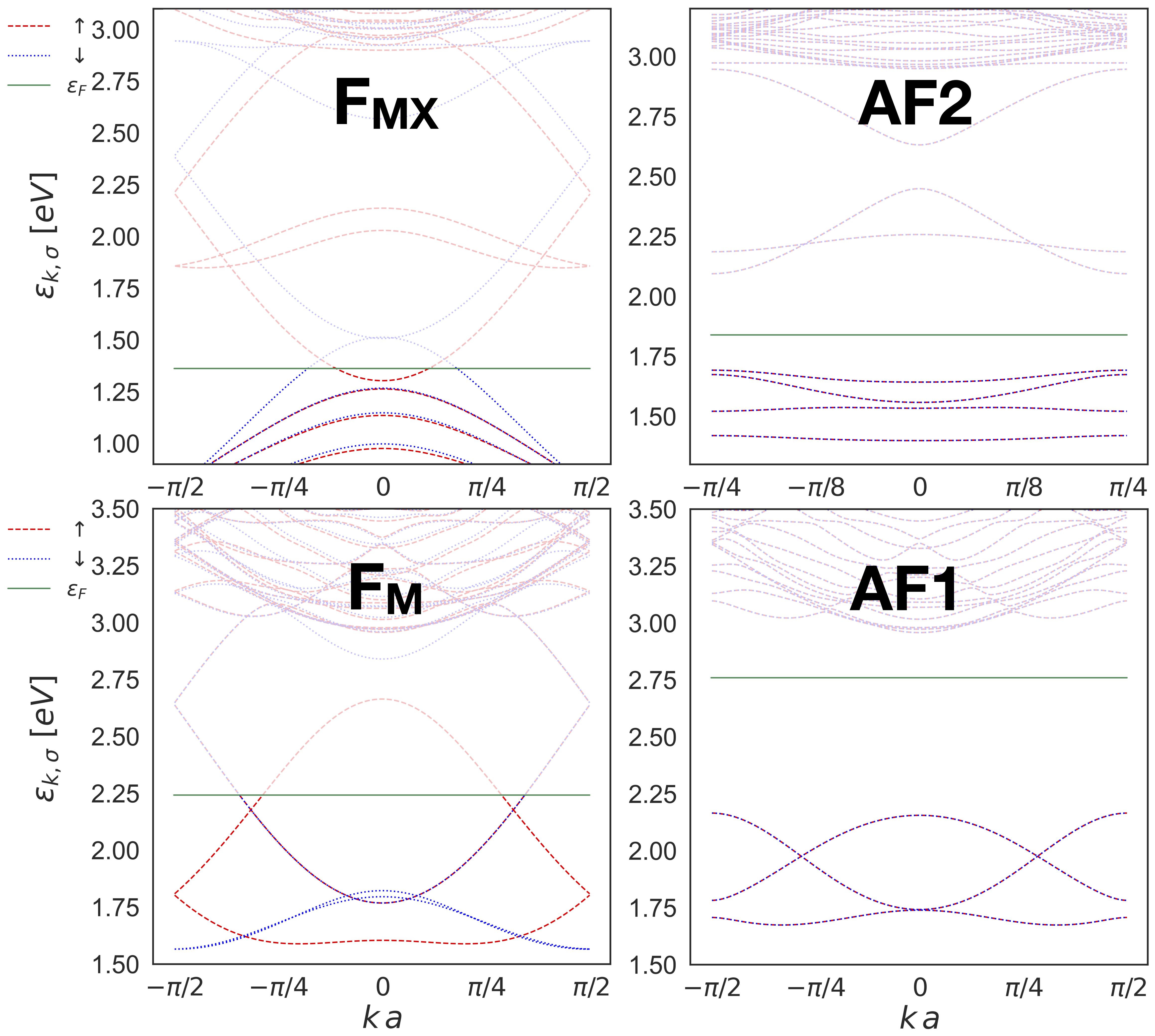}
\caption{Mean field band structure for edge fillings of $\nu_{edge}=0.23$
(top left - $\text{{F}}_{\text{{MX}}}$), $\nu_{edge}=0.5$ (AF2), $\nu_{edge}=0.60$ ($\text{{F}}_{\text{{M}}}$) and $\nu_{edge}=0.75$ (AF1). We used $U=2.94\,\text{eV}$ and $T=0$,
and a ribbon width of 10 $M$ atoms.
Spin-up bands are in dashed-red and spin-down in dotted-blue. Unoccupied bands are faded and
the horizontal green line marks the Fermi energy.
Ferromagnetic phases (left panels) are characterized by spin-splitting of edge bands and present
edge-metallicity (absence of a gap). In the $\text{{F}}_{\text{{MX}}}$ phase both edges are polarized
while in the $\text{{F}}_{\text{{M}}}$ phase only the M-edge is magnetized. 
The antiferromagnetic phases AF1 and AF2 (right panels) are insulating.
\label{fig:mf_bands}}
\end{figure}

In Fig.~\ref{fig:mf_bands}, we show the mean field band structures
for the two gapped phases AF1 and AF2 and two representative edge-ferromagnetic
phases. Both AF1 and AF2 band structures show an interaction-induced band gap (right panels).
The AF2 phase (top right) occurs
at $\nu_{edge} = 1/2$ and, correspondingly, half of the
edge-bands are filled (in this case, since $p=4$, this corresponds
to 4 out of 8 spin-degenerate bands). The AF1 phase (bottom right) occurs at $\nu_{edge} = 3/4$.
Since now we take $p=2$, this corresponds to 3 out of 4 spin-degenerate edge bands. 
For the two representative
edge-ferromagnetic phases (left panels), we took $p=2$ to accommodate the possibility
of (anti)ferrimagnetic ordering, but we consistently obtained ferromagnetism
(other types of magnetic ordering or paramagnetism were also energetically
excluded). At $\nu_{edge}=0.23$ (top left), both edges become magnetized since the
spin-degeneracy of all edge bands is lifted. For $\nu_{edge} = 0.60$ (bottom left), 
the bands corresponding to the $X$ edge remain
spin-degenerate, unlike the ones corresponding to the $M$ edge. Thus, only
the M-edge becomes magnetized.

\begin{figure}[ht]
\includegraphics[trim={4cm 0 4cm 0}, width=8.2cm]{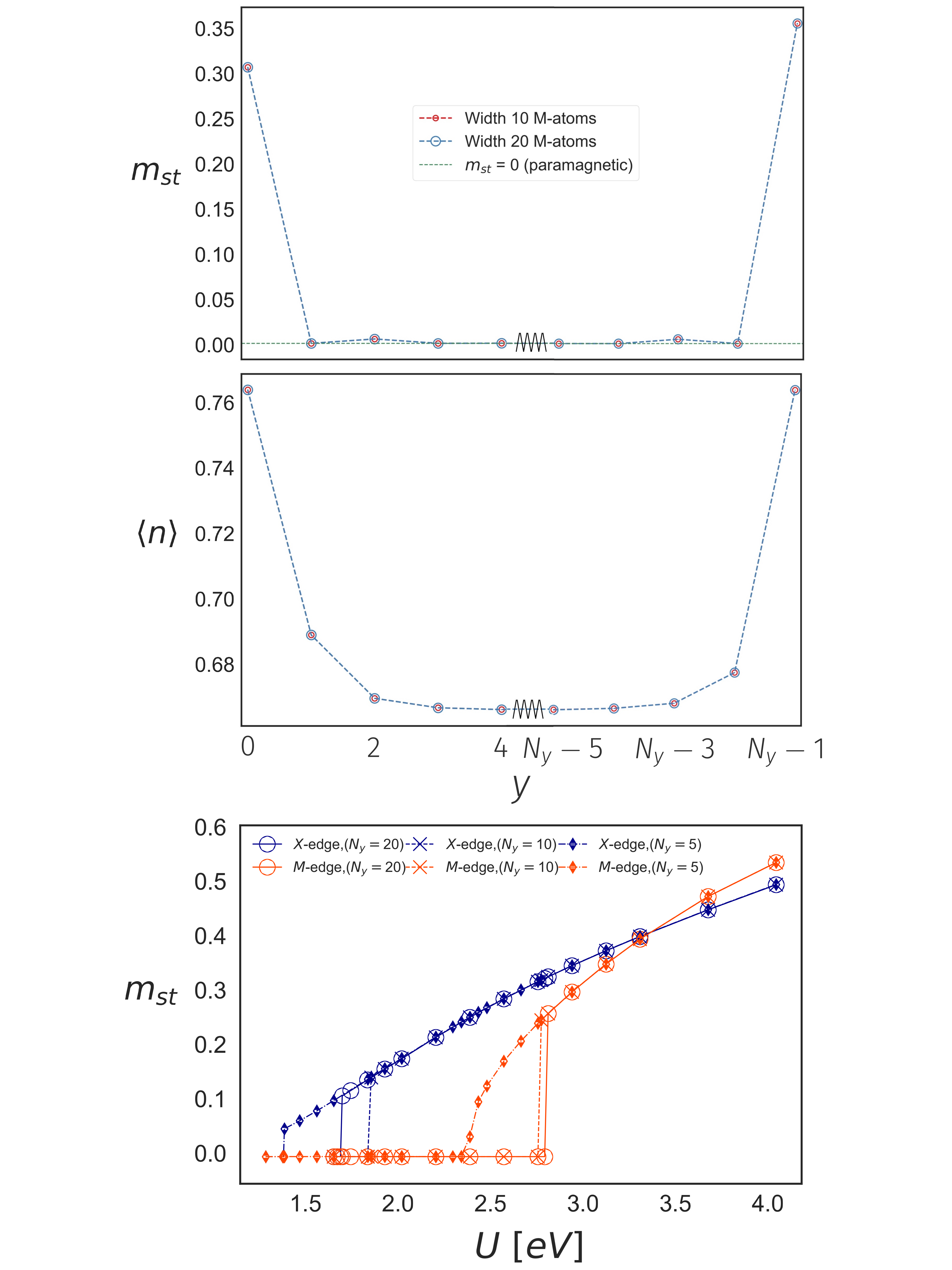}
\caption{Comparison between ribbons with widths $N_y=5, 10, 20$ $M$ atoms at $T=0$. (Top) Staggered magnetization $m_{st}$ and electron density $\left\langle n\right\rangle$ profiles as the row position $y$ is changed for $U=2.94\,\text{eV}$. 
Results for $N_y = 10$ and $N_y=20$ are indistinguishable.
(Bottom) $m_{st}$ vs $U$ at the X-edge (blue) and M-edge (orange).\label{fig:width-comparison}}
\end{figure}

The phases we have obtained are independent of the width of the  ribbon. To illustrate this, we
consider the AF1 phase at $\nu_{edge} = 3/4$. In Fig.~\ref{fig:width-comparison},
we show the staggered magnetization $m_{st}$ and the electron density
$\left\langle n \right\rangle$ (top and middle panels) as a function of the
row position $y$ for two different ribbon widths $N_y = 10$ and $N_y=20$ $M$ atoms. The results
are numerically indistinguishable. On the bottom panel of Fig.~\ref{fig:width-comparison},
we show the variation of $m_{st}$ with $U$ at $T=0$ for the
ribbon widths $N_y = 5, 10, 20$.

As pointed out earlier, edge-antiferromagnetism is more robust on the $X$ edge
for the three $N_y$ values considered, consistently appearing for lower critical values of $U$ compared to the M-edge.
The differences in the critical values of $U$ for different $N_y$ are related to the change
in $\left\langle n \right\rangle$
required to keep $\nu_{edge}$ fixed as $N_y$ is changed, according to Eq.~\eqref{eq:nu_edge}.
It is reasonable to assume
that systems with larger widths would behave similarly. In fact, for
the system with the smaller width of 5 $M$ atoms, the edge staggered magnetizations are
the same as those depicted on the top panel of Fig.~\ref{fig:width-comparison}. 
The fact that there
is no significant qualitative change justifies the use of a system
with $N_y=5$ $M$ atoms for the DetQMC calculations in Sec.~\ref{subsec:DetQMCres} (it becomes too computationally 
expensive to simulate larger systems using DetQMC due to the sign problem). 

To close this section, we note that the edge physics behind the magnetism we find in Fig.~\ref{fig:mf_phases} is the result of two competing mechanisms: gap opening instabilities and Stoner-like
edge-ferromagnetism. The winning mechanism depends on the edge filling, which is set via the Fermi energy. If the two Fermi points  of the noninteracting system are connected through a wave vector which spans an integer fraction of the Brillouin zone (dashed green and yellow lines of Fig.~\ref{fig:free-bands}), the addition of a mean field intraorbital Hubbard interaction induces a nesting instability which opens a gap (see right panels of Fig.~\ref{fig:mf_bands}). Other fillings favor Stoner-like
edge-ferromagnetism (see left panels of Fig.~\ref{fig:mf_bands}), with metallic edges and spin-split edge bands.

On the other hand, entropy gain due to thermal fluctuations tends to counteract magnetic ordering. This is illustrated in Fig.~\ref{fig:mf_phase_diagram}, where we observe that the critical Hubbard interaction required for magnetic ordering increases as the temperature is increased. 
The critical Hubbard interaction depends slightly on the width (see bottom panel of Fig.~\ref{fig:width-comparison}),  converging rapidly for wider ribbons.
Once the system becomes magnetic, the order parameters coincide regardless of the width (see top panel of Fig.~\ref{fig:width-comparison}), which is consistent with  edge-dominated physics.

\subsection{Mean field theory: multiorbital interaction}

\begin{figure}[ht]
\centering
\includegraphics[width=8.5cm]{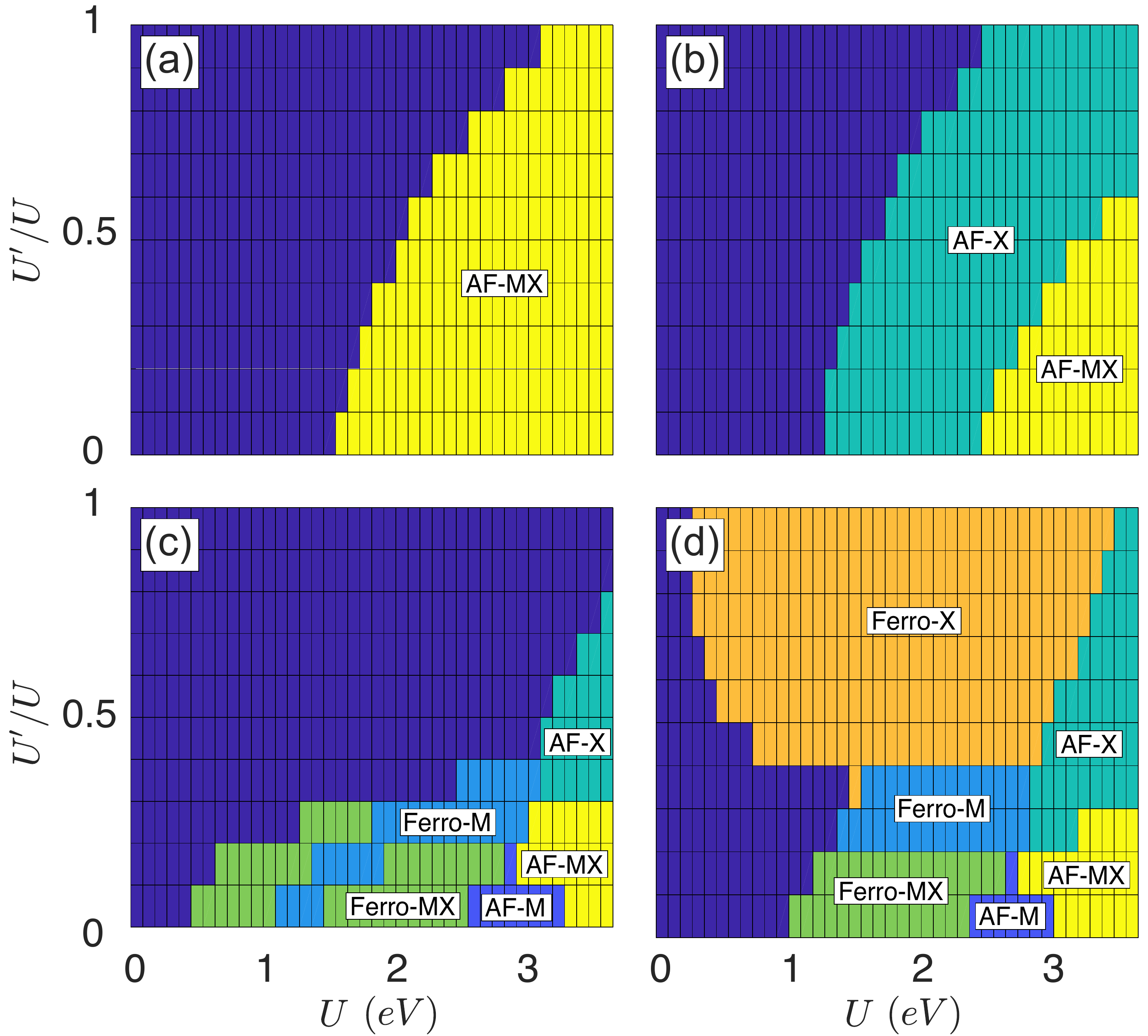}
\caption{$U^\prime{}-U$ phase diagrams at $T=0$ for the multiorbital case, obtained for a ribbon 
width $N_y=16$ at the edge fillings $\nu_{edge} = 75 \%$ (a)-(b), $\nu_{edge} = 62.5 \%$ (c), and
$\nu_{edge} = 65 \%$ (d). On panel (a) we consider $J= J^\prime{} = 0$ and
assume that both edges become polarized simultaneously. On the remaining panels, we have $J=J^\prime{} = (U-U^\prime{})/2$. Dark-blue regions are paramagnetic, and the labels in the
remaining phases stand for ferromagnetic on the X-edge (Ferro-X), on the M-edge (Ferro-M), or
on both edges (Ferro-MX), and antiferromagnetic on the X-edge (AF-X), on the M-edge (AF-M), or on
both edges (AF-MX). \label{fig:pd_interorbital}}
\end{figure}

In Fig.~\ref{fig:pd_interorbital}, we show the mean field phase diagrams for the multiorbital Hamiltonian, which includes both the minimal intraorbital model and the terms of Eqs.(\ref{eq:add_terms0}-\ref{eq:add_terms2}).
As stated in Sec.~\ref{subsec:mean field-theory}, we assume a smaller set of mean field parameters 
by assuming that only the edges get magnetized. This is a justified approach based on the results of the 
previous section (in particular, the top panel of Fig.~\ref{fig:width-comparison}). The results obtained
with multiorbital interactions are compiled in Fig.~\ref{fig:pd_interorbital}, where we show phase
diagrams in the plane $U$ versus $U^\prime{}$ at $T=0$, obtained with a ribbon of width $N_y = 16$.

We start by focusing on the AF1 phase at $\nu_{edge} = 3/4$.
In Fig.~\ref{fig:pd_interorbital}(a) we consider $J=J^\prime{}=0$ and assume both edges are
simultaneously magnetized. It is clear that the interorbital term $U^\prime{}$ counteracts the tendency for antiferromagnetic order since the critical $U$ value required for the onset of AF1 phase increases as $U^\prime{}$ is increased. Figure~\ref{fig:pd_interorbital}(b) shows the effect of including $J$ and
$J^\prime{}$, now allowing for edge magnetization independently on each edge. According to Sec.~\ref{subsec:mean field-theory}, we consider $J=J^\prime{}=(U-U^\prime{})/2$. It can be seen that
the X-edge becomes polarized first, in agreement with the results of the previous section.
When $U^\prime{}=0$ and $J=J^\prime{}=U/2$, 
one can see that the critical $U$ for the onset of X-edge magnetization (AF-X), as well as for the onset 
of magnetization on both edges (AF-MX), is lower than the case when $U^\prime{}=J=J^\prime{}=0$ shown in
Fig.~\ref{fig:width-comparison} (bottom panel). This indicates that $J$ and $J^\prime{}$ favor
magnetic order.
Similarly to the case of panel~\ref{fig:pd_interorbital}(a), as $U^\prime{}$ increases, the critical value of $U$ for the onset of antiferromagnetism increases. 

We now consider two representative edge fillings between $\nu_{edge}=1/2$ and $\nu_{edge}=3/4$. 
In Fig.~\ref{fig:pd_interorbital}(c) we set $\nu_{edge} = 62.5 \%$. Two ferromagnetic phases can
be seen, one on both edges (Ferro-MX) and the other solely on the M-edge (Ferro-M).
Even though ferromagnetic phases appear for lower values of $U$, antiferromagnetic phases
eventually appear as $U$ is increased. Yet, the interorbital interaction $U^\prime{}$ still suppresses magnetic ordering quite significantly. The edge filling  $\nu_{edge}=65 \%$ is
considered in Fig.~\ref{fig:pd_interorbital}(d).
Ferromagnetism becomes more prominent on the phase diagram, with a ferromagnetic phase on the X-edge (Ferro-X) appearing. Moreover, the tendency for the interorbital interaction to suppress magnetic 
ordering is weakened, with the Ferro-X and AF-X phases still surviving even as $U^\prime{}$ approaches $U$. Notably, the Ferro-X phase becomes favorable for lower values of $U$ as $U^\prime{}$ is increased.

\subsection{Determinant Quantum Monte Carlo\label{subsec:DetQMCres}}

We now turn to the DetQMC approach (see Appendix \ref{sec:B} for specific details about our implementation). For our minimal Hubbard model, the method is severely limited
by the sign problem, with the average sign going to zero in most regions
of interest of the phase diagram. Notwithstanding, we are able to
confirm the appearance of the AF1 phase predicted with MFT.

\begin{figure}[ht]
\centering\includegraphics[width=8.2cm]{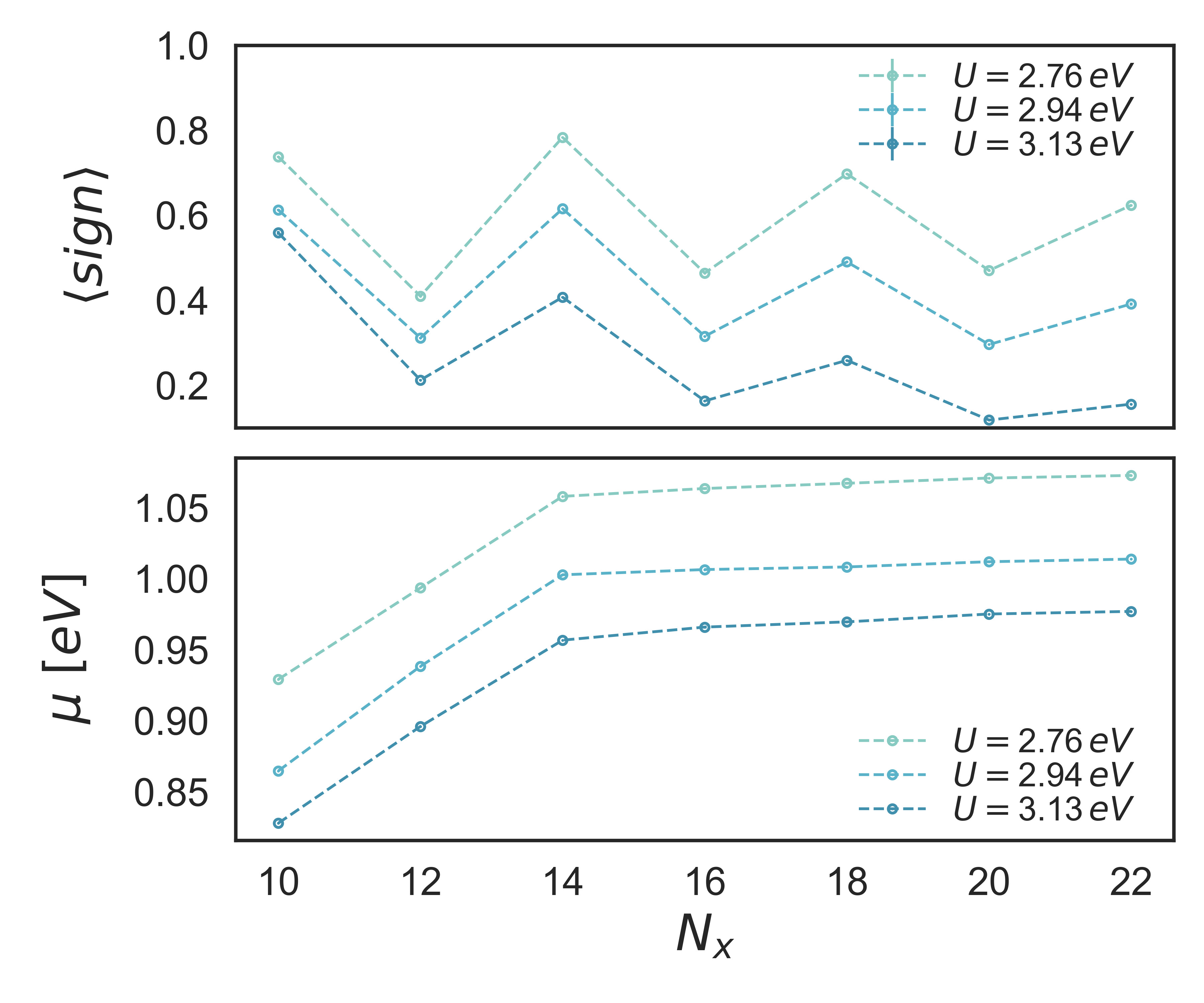}
\caption{(Top) Average sign obtained in the Monte Carlo sampling
for varying $N_{x}$ and $U$ at fixed temperature $T=267\,\text{K}$, for $\text{{Mo}\text{{S}}}_{2}$
nanoribbons of 5 $M$ atoms of width for electron densities corresponding
to $\nu_{edge}\approx0.75$ (error bars are negligibly small). (Bottom)
Chemical potential used in the DetQMC algorithm to obtain the required
edge filling for each system size (other parameters are kept the same as
on the top panel). \label{fig:qmc_sign}}
\end{figure}

On the top panel of Fig.~\ref{fig:qmc_sign}, we show the average sign for some of the 
parameters ($N_x, U$) we use throughout this section. The width is $N_y = 5$  $M$ atoms.
According to Eq.~(\ref{eq:nu_edge}), the electron density corresponding to $\nu_{edge}=3/4$ is $\left\langle n\right\rangle = 11 / 15$. For each system size $N_x$ in the longitudinal direction, we fix the chemical potential so as to approximate this electron density, measured with DetQMC, as closely as possible.
On the bottom panel of Fig.~\ref{fig:qmc_sign}, we show the chemical potential
required in order to obtain $\left\langle n\right\rangle _{QMC}\approx 11 / 15$.
Note that it initially grows with the system size, but then tends to stabilize. This is already apparent for $N_{x}=16$.

\begin{figure}[ht]
\includegraphics[trim={0 0cm 2.15cm 1.75cm},width=8cm]{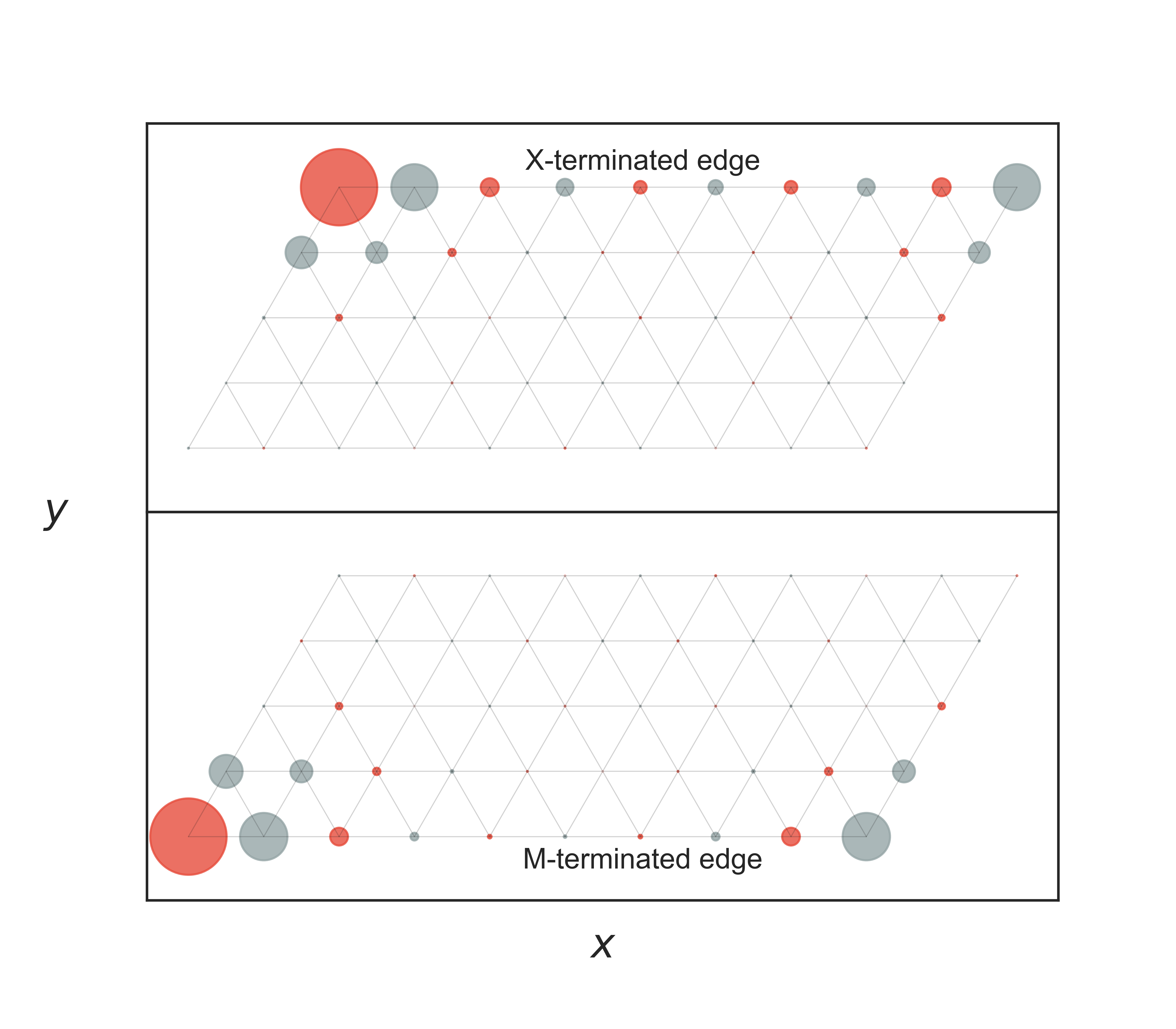}
\includegraphics[trim={0 1cm 0 0},width=8.5cm]{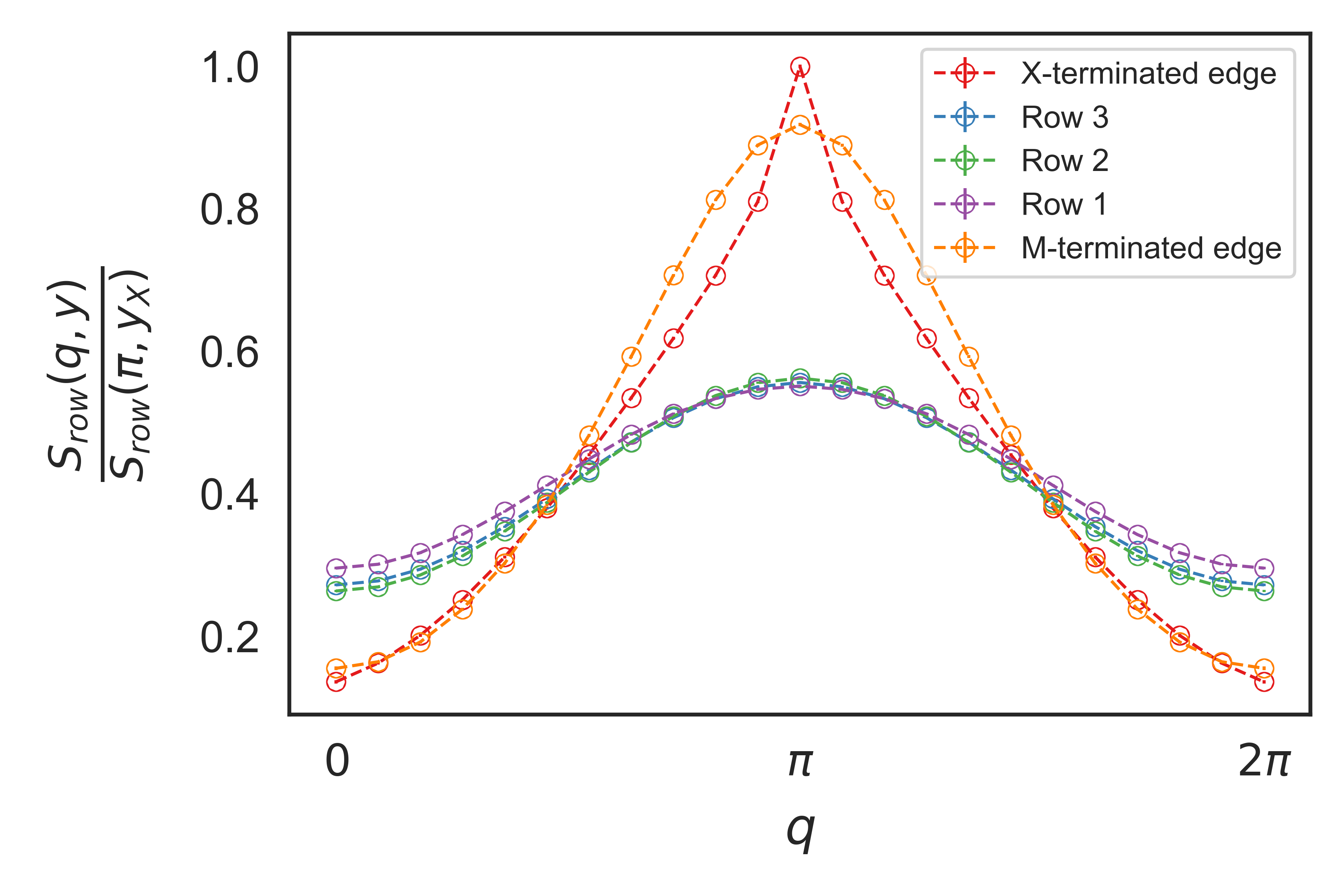}
\caption{Evidence for the AF1 phase obtained with DetQMC for $\text{{Mo}\text{{S}}}_{2}$ nanoribbons, with $U=2.76\,\text{eV}$ and $T=267\,\text{K}$. /Top) Spin-spin correlations for a $10\times5$ ribbon for edge filling $\nu_{edge}=0.75\pm0.01$, measured with respect to the leftmost site of the X(M)-edge.
The size of the circles indicates the magnitude of the correlations (stronger on the X-edge). Red corresponds
to a positive correlation and blue to a negative correlation.
(Bottom) Magnetic structure factor per row, normalized to its maximum value
($q=\pi$, at the X-edge). Here, we consider a
$22\times5$ $\text{{Mo}\text{{S}}}_{2}$ ribbon with $\nu_{edge}=0.745\pm0.008$. The error
bars are negligibly small. \label{fig:Spin-spin-correlations}}
\end{figure}

On the top panel of Fig.~\ref{fig:Spin-spin-correlations}, we show the spin-spin correlator along the edges
measured with DetQMC.
For $\nu_{edge}\approx 0.75$ --- for which the AF1 phase appears in MFT --- our DetQMC
results show that the spin-spin correlator has an alternating pattern that signals
antiferromagnetic ordering. The staggered pattern corresponds to a peak at $\pi$ in $S_{row}(q, y)$ computed for the edges of the ribbon ($y=0, y=N_y-1$), shown on the bottom panel of Fig.~\ref{fig:Spin-spin-correlations}. This peak is considerably more pronounced on the edges than on the other rows of the ribbon, indicating a tendency towards edge-antiferromagnetic ordering. The sharper peak for the X-edge compared to the M-edge confirms that antiferromagnetism is more robust on the former.

\begin{figure}[ht]
\includegraphics[width=7.8cm]{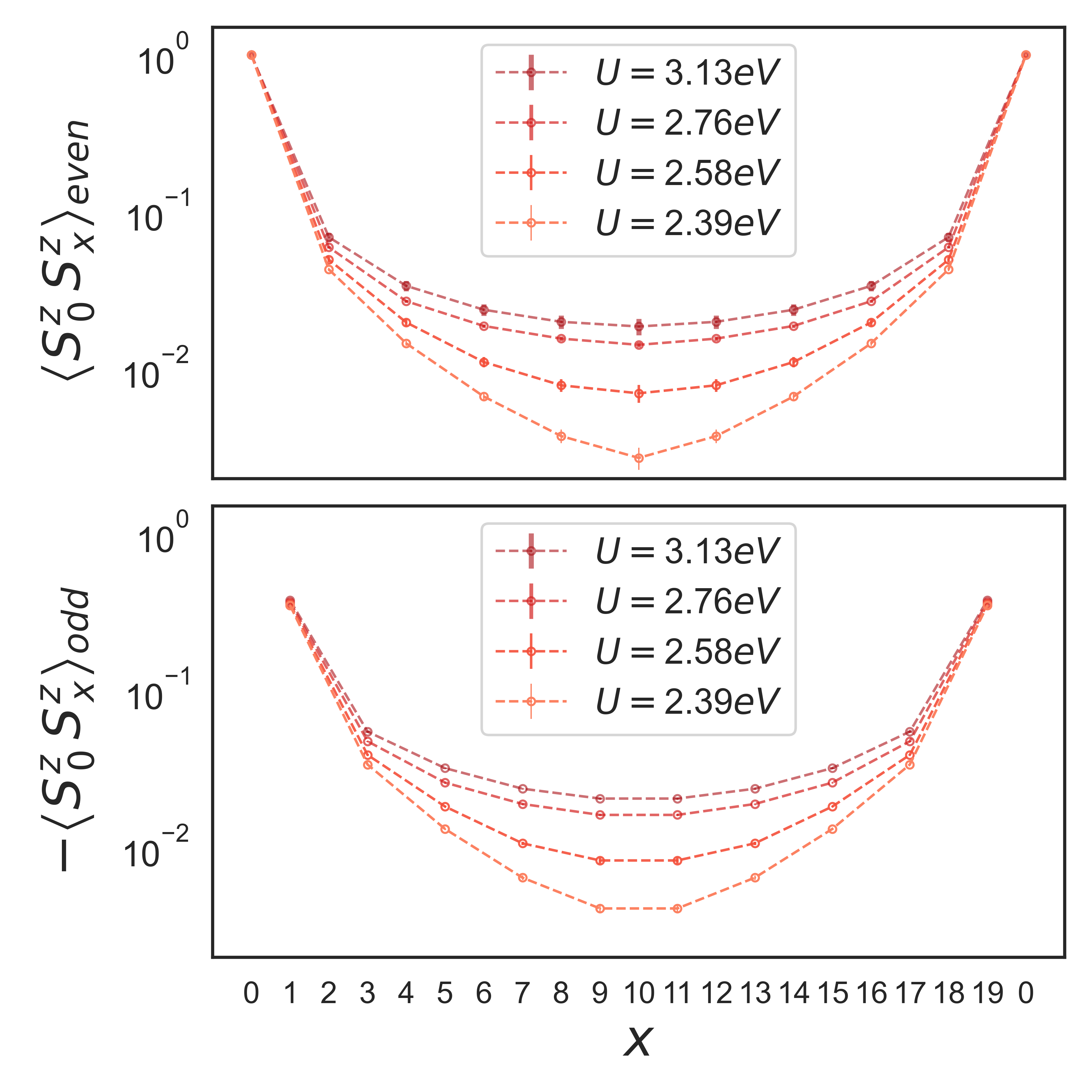}
\includegraphics[width=7.8cm]{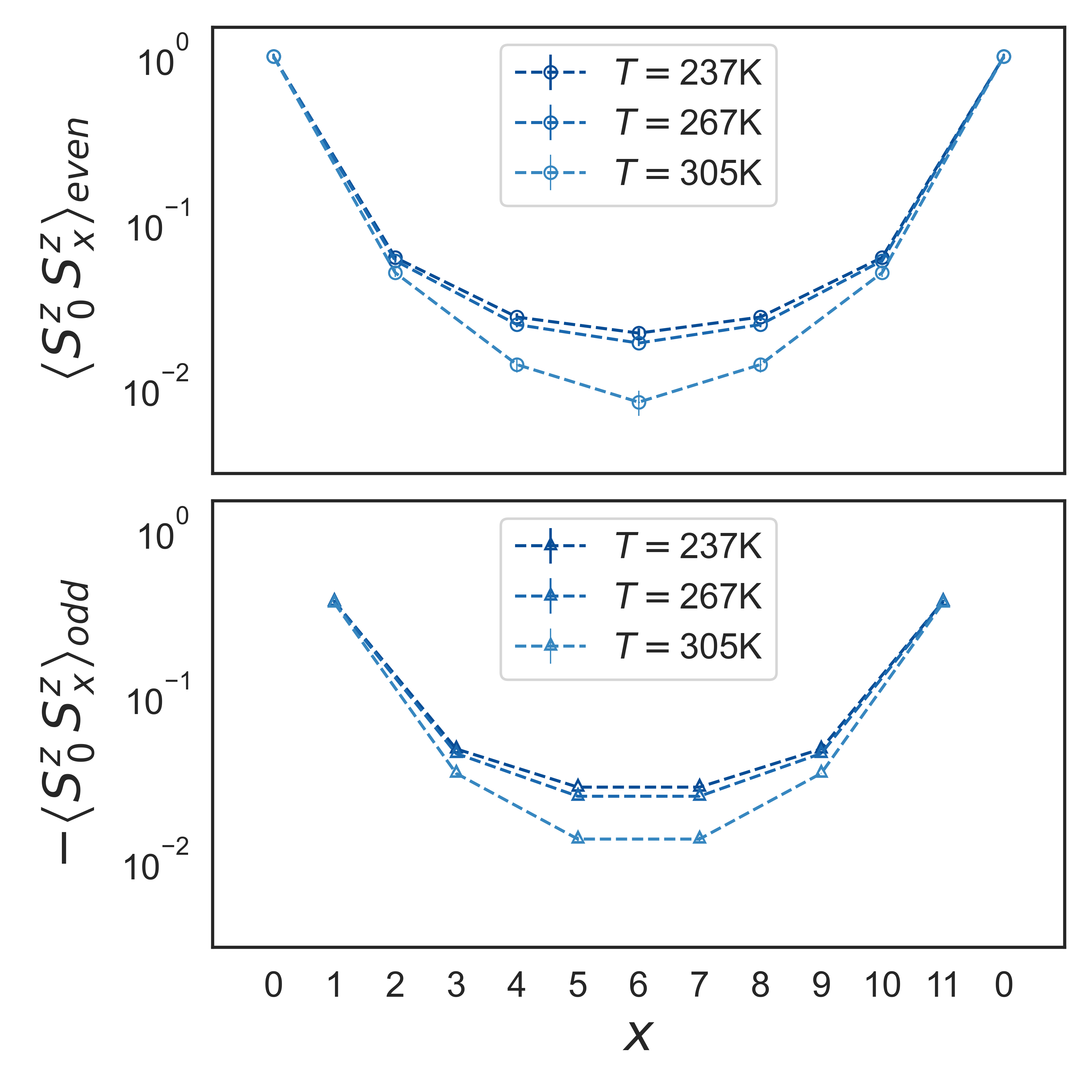}
\caption{Spin-spin correlators between even and odd  sites, along
the X-edge of $\text{{Mo}\text{{S}}}_{2}$ nanoribbons measured with DetQMC for $\nu_{edge}\approx 0.75$. On the top panels, we vary the Hubbard interaction $U$ and fix the temperature
$T=267\,\text{K}$ for a $20\times5$ ribbon. On the bottom panels, we vary $T$ and fix 
$U=2.94\,\text{eV}$ for a $12\times5$ ribbon.
\label{fig:Spin-spin-correlators-U-beta}}
\end{figure}

In Fig.~\ref{fig:Spin-spin-correlators-U-beta}, we show that as the Hubbard interaction $U$ (top panels) or the inverse temperature $\beta$ (bottom panels) increase, the spin-spin correlations at the X-edge increase in magnitude. As can be seen on the bottom panels, for $T=267\,\text{K}$ the behavior of the spin-spin correlations does not differ significantly from those of the system at lower temperature $T=237\,\text{K}$.
As the temperature increases, the average sign gets closer to 1, yielding less statistical
fluctuations. Thus, in the remainder of this section, we fix $T=267\,\text{K}$ since it  gives statistically relevant results which are already sufficiently close to the zero temperature limit. Note that, taking the semiconducting bulk gap $\Delta \sim 1\,\text{eV}$ in 2D TMDs as an estimate for the bandwidth of edge states, we obtain $\beta \Delta \sim 43$ at $T=267K$, which is well within typical values used  to simulate ground state properties with DetQMC.

\begin{figure}[ht]
\hspace{-6.38mm}\includegraphics[width=8.52cm]{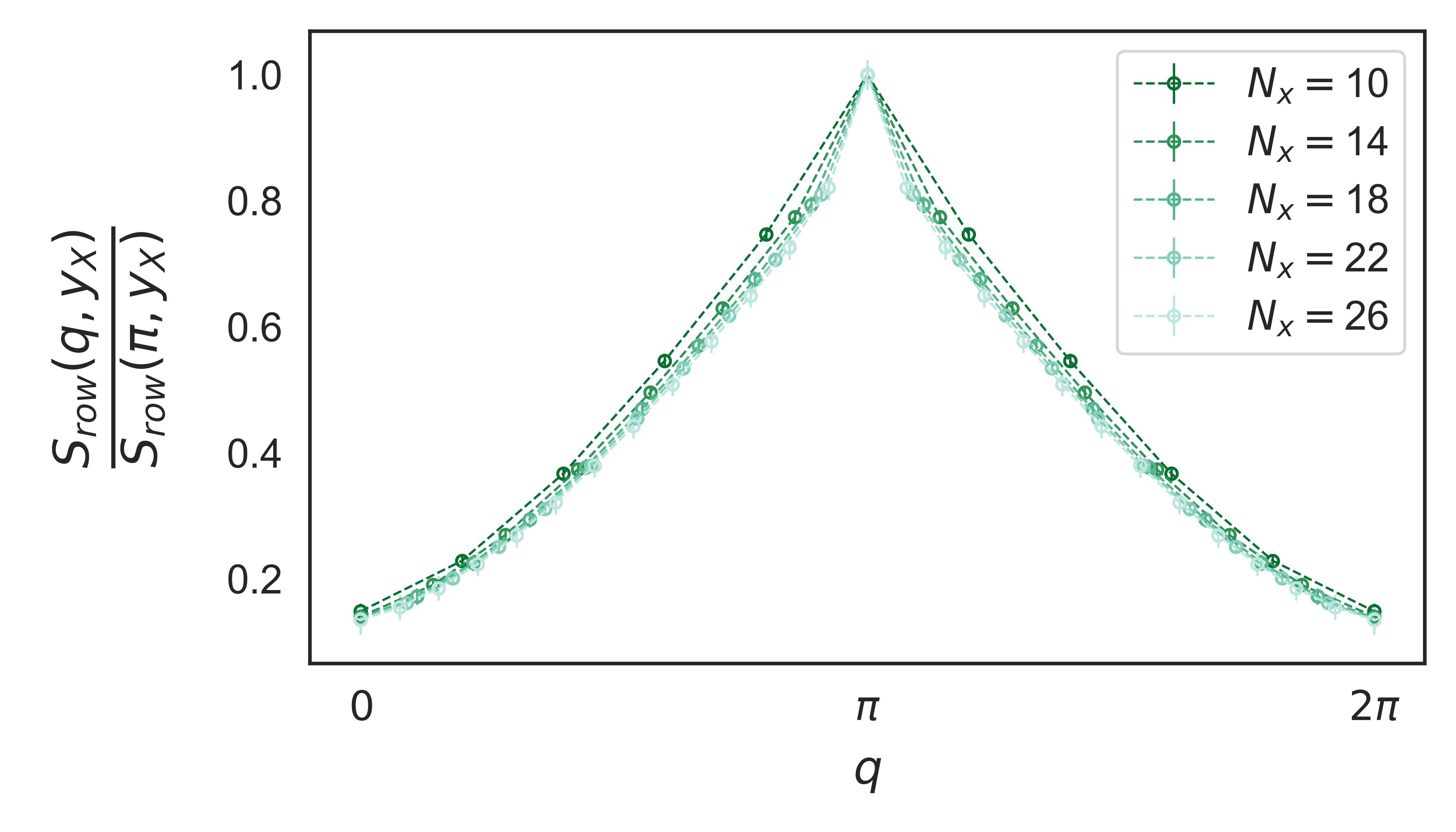}
\includegraphics[trim=8mm 0 1cm 0,width=7.9cm]{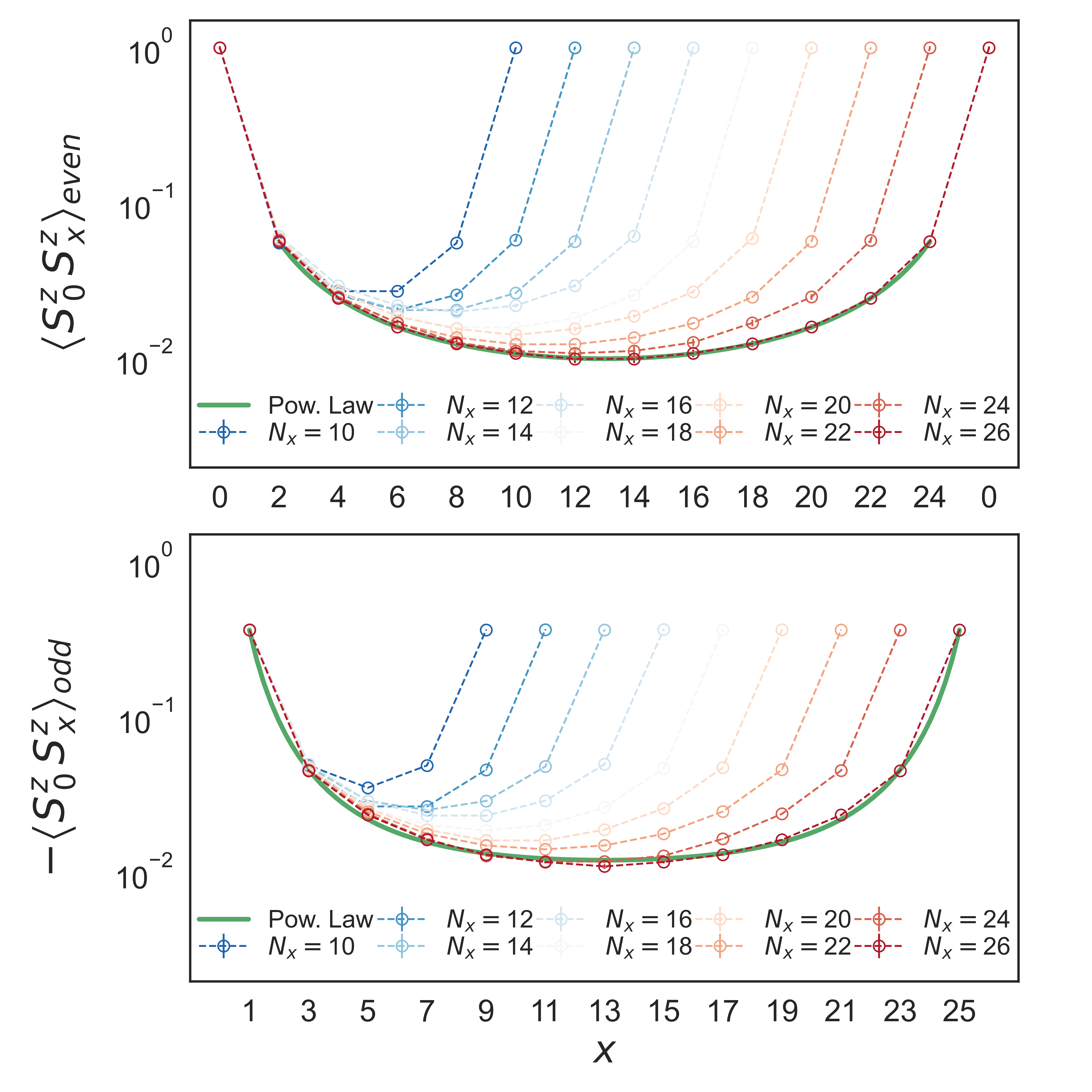}

\caption{(Top) Magnetic structure factor $S_{row}(q, y_X)$ normalized to the peak  $S_{row}(\pi, y_X)$ for nanoribbons with a width $N_y=5$ $M$ atoms and $U=2.76\,\text{eV}$ at $T=267\,\text{K}$ for $\nu_{edge}\approx0.75$. The $q=\pi$ peak sharpens as $N_x$ increases.
(Bottom) Spin-spin correlations
corresponding to the curves on the top panel with additional data shown. We use $N_{x}=10,...,26$ in steps of 2. The error bars are smaller than the symbols. The green curves are fits to the DetQMC data using functions that decrease as a power law. \label{fig:corrVar}}
\end{figure}

By varying the longitudinal dimension
of the ribbon, $N_{x}$, we are able to extrapolate the value of the
staggered magnetization to the thermodynamic limit using the method
outlined in Sec.~\ref{subsec:DetQMC}.
In Fig.~\ref{fig:corrVar}, we show finite-size scaling data for $U=2.76\,\text{eV}$.
On the top panel, one can see that the $q=\pi$ peak of $S_{row}(q, y_X)$ sharpens as $N_{x}$ increases. On the
bottom panel, it is seen that the staggered spin-spin correlation on the X- edge tends to a constant as the system size is increased, which signals antiferromagnetic ordering.

We find that the spin-spin correlations decay algebraically, indicating quasi long-range order.
To fit the results, we use the power law:
\begin{equation}
(-1)^x\left\langle S^z_{\,0} S^z_{\,x} \right\rangle = \bigg(\frac{x}{\xi}\bigg)^{1-\eta} + \bigg(\frac{N_x -x}{ \xi}\bigg)^{1-\eta} + \text{constant},
\end{equation}
where $\xi$ and $\eta$ are respectively the correlation length and the critical exponent, with $\xi= \xi_\text{even}, \xi_\text{odd}$ and $\eta = \eta_\text{even}, \eta_\text{odd}$ depending on whether $x$ is odd or even.
By fitting to the DetQMC data for $N_x = 26$, we find the correlation lengths: $\xi_\text{even} = 0.225 \pm 0.003$ and $\xi_\text{odd} = 0.605 \pm 0.003$. These are consistent with our scaling hypothesis in Eq.~\eqref{eq:scaling_form} since $\xi_\text{even/odd} \ll N_x$. We also find the critical exponents $\eta_\text{even} = 2.343\pm0.008$ and $\eta_\text{odd} = 3.06\pm0.02$.

\subsection{Comparison between MFT and DetQMC}

\begin{figure}[ht]
\hspace{-1mm}\includegraphics[width=7.9cm]{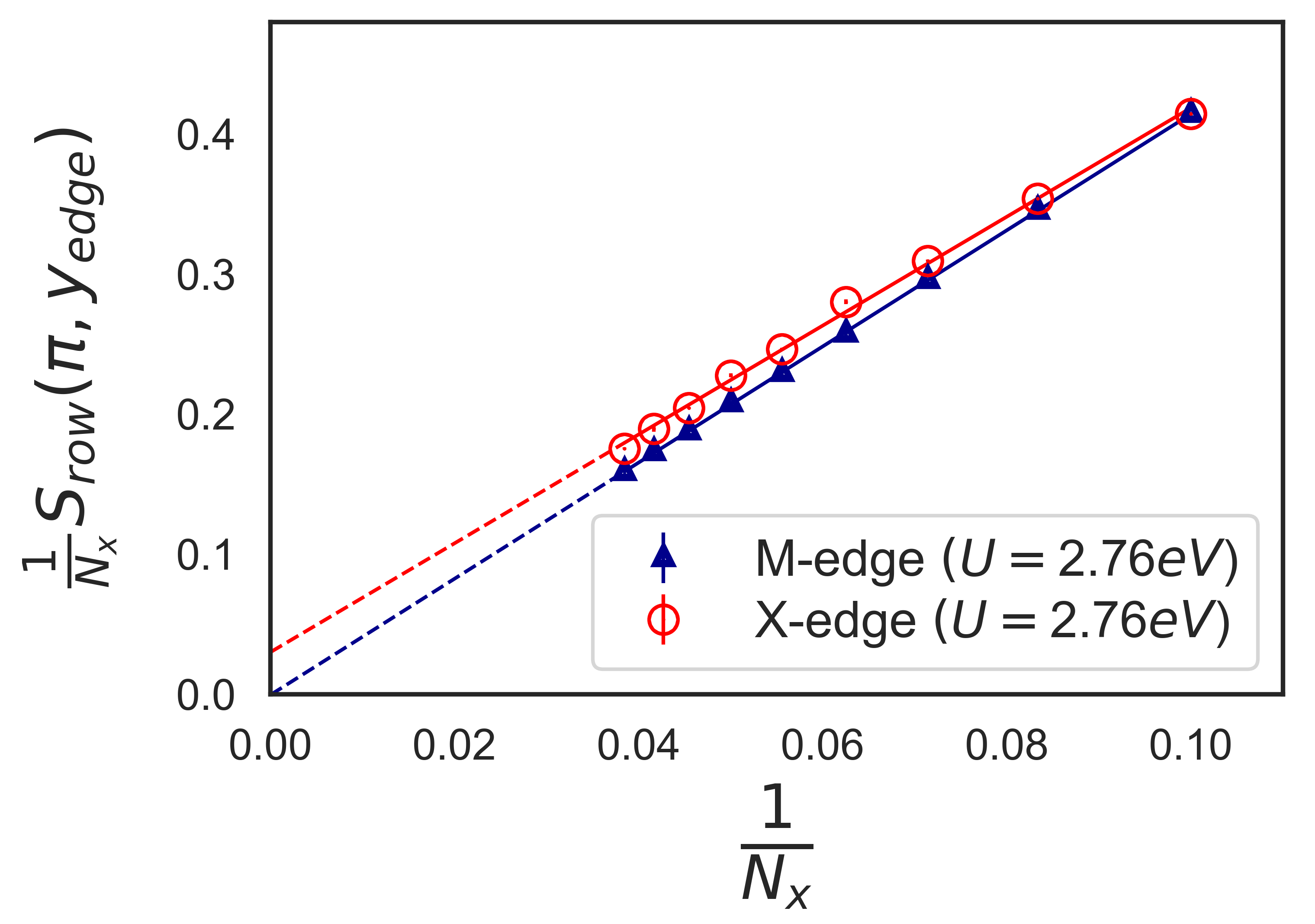}
\includegraphics[width=7.8cm]{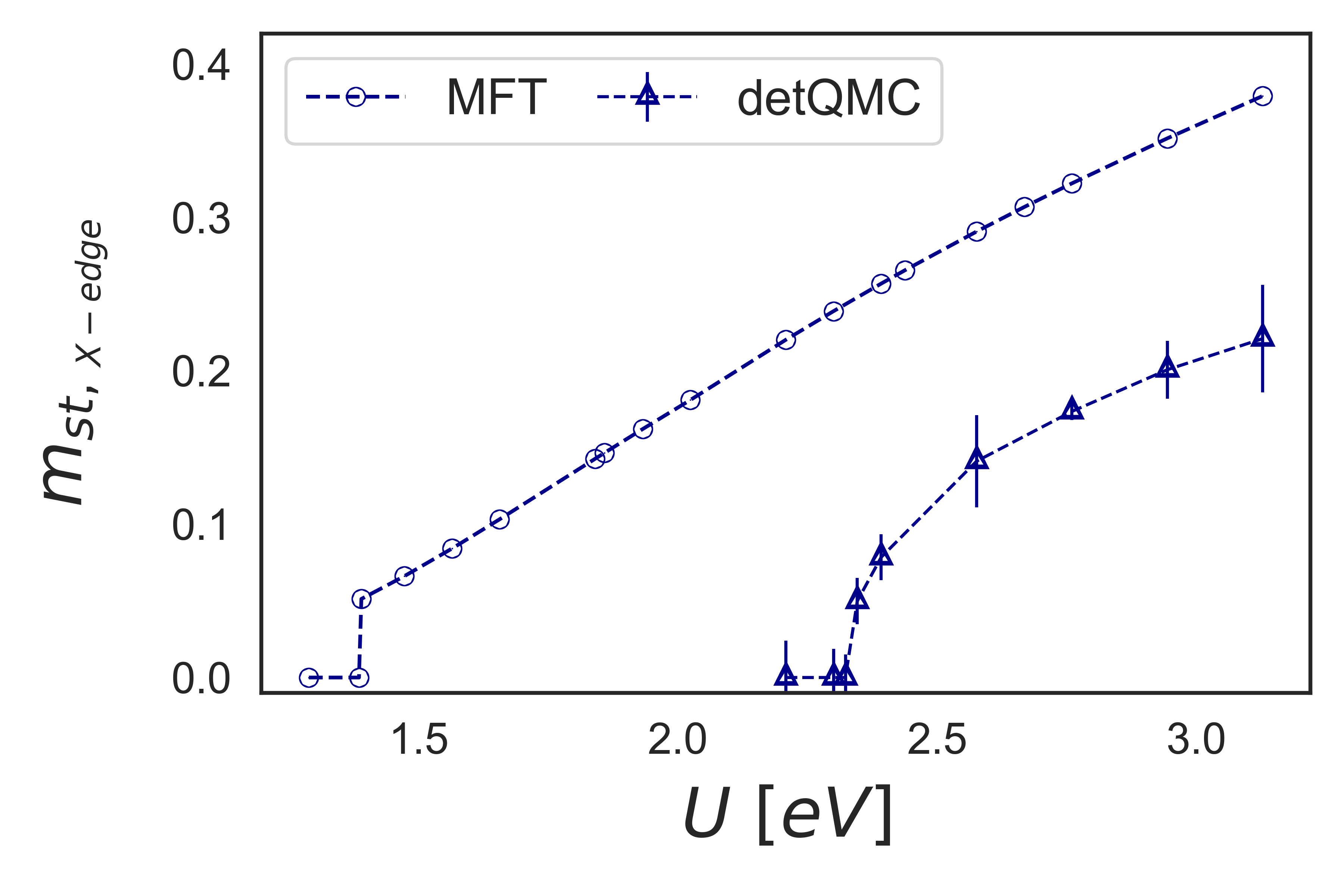}
\caption{(Top) DetQMC results for $S_{row}(\pi, y)/N_x$, with  $y=y_{edge}$, as a function of $1/N_x$, for  nanoribbons with a width of $N_y =5$ $M$ atoms at $T=267\,\text{K}$ and $\nu_{edge}\approx 3/4$. Red circles are for X-edge and blue triangles for M-edge.
The lines are fits to the data using Eq.~\eqref{eq:extrapol}. Dashed lines are extrapolations to the  thermodynamic limit (we repeat the extrapolation illustrated on this panel for varying Hubbard interaction $U$ to produce the bottom panel). (Bottom) Comparison between the staggered magnetizations on the X-edge obtained with MFT (circles) and DetQMC (triangles) as a function of $U$. MFT results are for $N_y=5$ and $T=0$.\label{fig:mf_qmc}}
\end{figure}

We close this section with a critical comparison of the results obtained from MFT and DetQMC at $\nu_{edge} = 3/4$, where the AF1 phase shows up. In order to obtain the staggered magnetization based on the DetQMC results, we plot $S_{row}(\pi, y)/N_x$ for $y=y_{edge}$ as a function of $1/N_{x}$, and use Eq.~\eqref{eq:extrapol} to extrapolate to the thermodynamic limit. Representative results are shown on the top panel of Fig.~\ref{fig:mf_qmc}.

Notice that although we observe antiferromagnetic correlations at the M-edge with DetQMC, these do not signal magnetic ordering in the thermodynamic limit: the blue dashed line on the top panel intersects the $y$-axis approximately at zero. It is possible that antiferromagnetism on the M-edge appears for higher values of $U$ similarly to what we observed with MFT. However, we cannot confirm this suspicion because the sign problem becomes too severe beyond $U=3.13\,\text{eV}$ (at which point our DetQMC simulations show no sign of antiferromagnetic order on the M-edge).

On the bottom panel of Fig.~\ref{fig:mf_qmc}, we compare the extrapolated staggered
magnetization obtained with DetQMC  to the MFT results. We considered
the same ribbon width $N_y=5$ in MFT but used $T=0$. 
As expected, DetQMC predicts  a higher critical interaction, $U_{c,QMC}\approx2.33\pm 0.02\,\text{eV}$,  compared to the MFT result, $U_{c,MFT}=1.387\pm0.004\,\text{eV}$.
The agreement becomes better if we consider wider ribbons and/or $T>0$ in the MFT 
calculation.
Qualitatively, the MFT and DetQMC results are similar
in the sense that edge-antiferromagnetic ordering is established for values of the interaction that are of the same order of the band gap $U_c \sim \Delta$, with MFT overestimating long-range ordering. DetQMC unveils quasi-long-range order, with algebraic behavior of the spin-spin correlations.

\section{Discussion and conclusions\label{sec:conclusions}}

We have used MFT and DetQMC to probe edge magnetism in zTMDNRs via a minimal three-band Hubbard model. Three main questions have been addressed: (1) How does changing the edge filling affect the phase diagram? (2) What is the effect of multiorbital interactions? (3) Can we use the numerically exact DetQMC approach to probe edge magnetism in TMD nanoribbons in spite of the well known sign problem?

To answer the first question, we considered an intraorbital Hubbard $U$ interaction, which we treated at the mean field level. We found two gapped phases: an edge-dimer phase (AF2), when the edges are half filled, and an edge-antiferromagnetic
phase (AF1) at three-quarter edge filling. For other edge fillings, there is a tendency towards edge-ferromagnetism. As shown in Fig.~\ref{fig:mf_bands}, the ferromagnetic phases are metallic and the edge magnetization depends on the edge filling. Such magnetic edge states  give rise to spin-polarized edge currents which could be tuned by changing the Fermi level through a back gate voltage. In particular, when the gapped phases are reached, these currents are suppressed.  Similar behavior has been explored in zTMDNRs in the presence of magnetic proximity effect produced by ferromagnetic \cite{cortes_tunable_2019,Anisotropy2019LChico} and antiferromagnetic \cite{Cortes2020} substrates. Our results indicate that intrinsic magnetism could also be used to induce spin-polarized edge currents.

The second question has been answered by considering not only the intraorbital Hubbard $U$ interaction, but also an interorbital interaction term $U^\prime{}$, as well as Hund $J$ and pair-hopping $J^\prime{}$ terms, characteristic of transition metal atoms. Within MFT, we obtained rich phase diagrams shown in Fig.~\ref{fig:pd_interorbital}, which corroborate and further complement the phases obtained with the simpler intraorbital Hubbard $U$ interaction. Generically, the interorbital $U^\prime{}$ term tends to suppress the magnetic phases, while $J$ and $J^\prime{}$ tend to enhance them. However, there are fillings for which $U^\prime{}$ stabilizes the ferromagnetic phase in a large portion of the phase diagram, as shown in Fig.~\ref{fig:pd_interorbital}(d).

As far as the third question is concerned, we have successfully applied DetQMC to the intraorbital Hubbard $U$  model for zTMDNRs, finding that edge magnetism strongly depends on the edge filling. In particular, at three-quarter edge filling -- where the AF1 phase appears at mean field level -- DetQMC has only a moderate sign problem and accurate results can be obtained. We have found edge-antiferromagnetic quasi-long-range order with spin-spin correlations behaving algebraically, reinforcing the AF1 phase predicted by MFT. The extrapolated staggered magnetization from DetQMC is consistent with the MFT result, as shown in Fig.~\ref{fig:mf_qmc}. Even though in MFT long-range
order is slightly overestimated, DetQMC and MFT agree that antiferromagnetism is more robust on the X-edge.

Let us point out that the dependence of edge magnetism on edge filling that we found might be relevant when interpreting experimental results. Often, the density of zigzag edges is used to explain how the ferromagnetic response varies between different nanosheet samples. Our results point to the edge filling as yet another key ingredient, since structural defects or chemisorbed adatoms may effectively change the filling of the edge. Finally, an important aspect for further study is the impact of a magnetic substrate --- which induces magnetic exchange fields as considered in Refs.~\cite{cortes_tunable_2019,Anisotropy2019LChico,Cortes2020} --- on the edge magnetism of zTMDNRs. The methods we use in this work could be used to determine whether
the edge magnetism we have found survives the presence of
a substrate, and whether phase transitions can be induced by tuning
the coupling to the substrate. Searching for edge magnetism at twin grain boundaries in 2D TMDs \cite{Cadez2019} and at 1D interface-states in TMD heterostructures \cite{Avalos-Ovando2019} is another interesting direction.

\section{acknowledgements}

We acknowledge support by the Portuguese Foundation for Science and Technology through Strategic Funding No. UIDB/04650/2020, Project No. POCI-01-0145-FEDER-028887. F.M.O.B. is supported by a DTP studentship funded by the Engineering and Physical Sciences Research Council. The authors would like to gratefully acknowledge João Pedro dos Santos Pires and Simão Meneses João for the enlightening discussions. This project was undertaken on the Viking Cluster, which is a high-performance computing facility provided by the University of York. We are grateful for computational support from the University of York High Performance Computing service (Viking) and the Research Computing team.

\appendix
\section{Effect of edge bandwidth on edge magnetism \label{sec:A}}

In this appendix, we discuss the impact of using different TMDs in our calculations, therefore changing the edge bandwidth, i.e., the portion of the spectrum corresponding to edge bands. Using different TMDs has only a slight impact on the edge magnetic ordering because the edge bandwidth is similar ($\sim 1 $ eV) among group 6 TMDs, as we show in Fig.~\ref{fig:Band-structures-TMDNR}.

\begin{figure}[ht]
\centering\includegraphics[width = 8. cm]{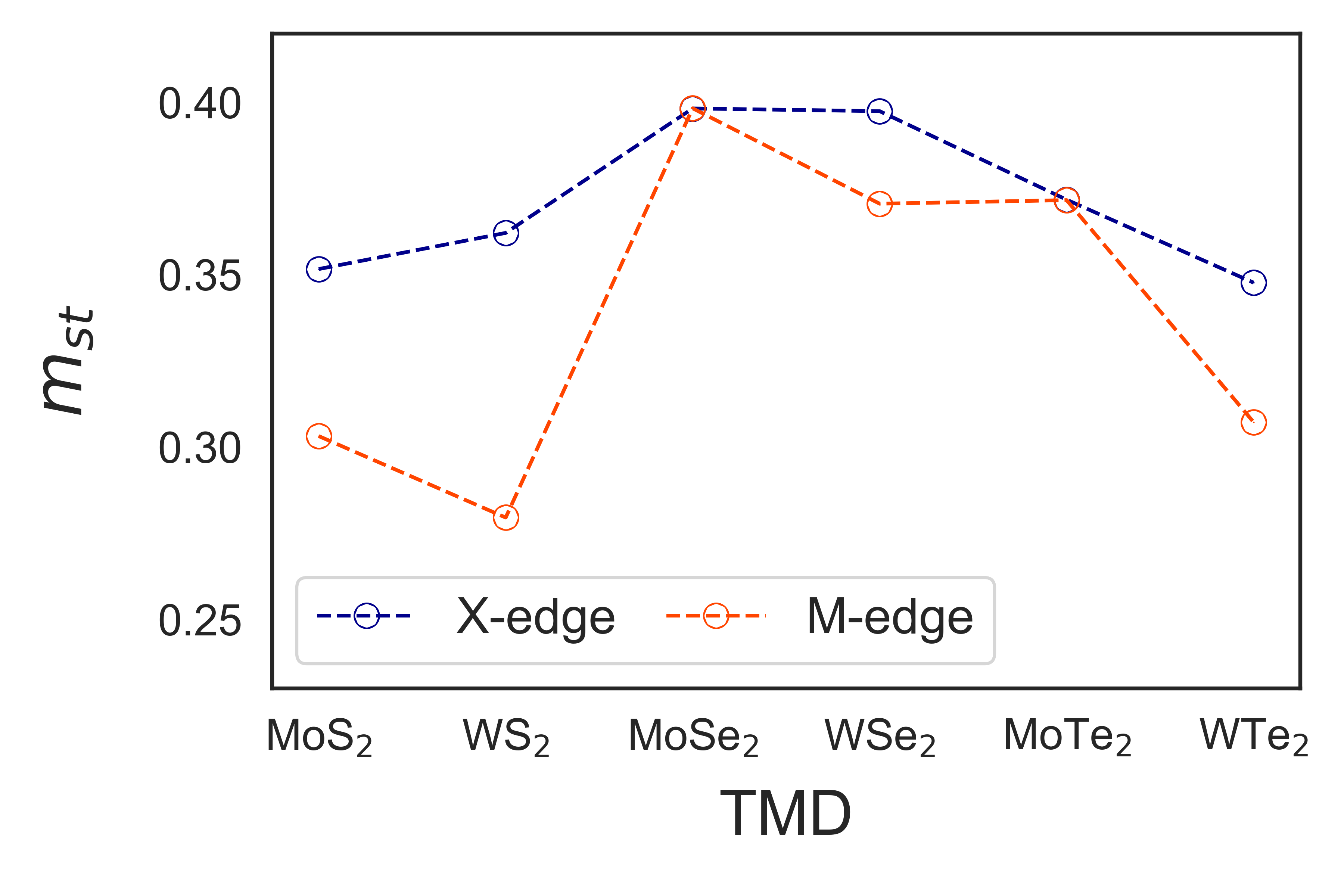}
\smallskip
\includegraphics[width = 3cm]{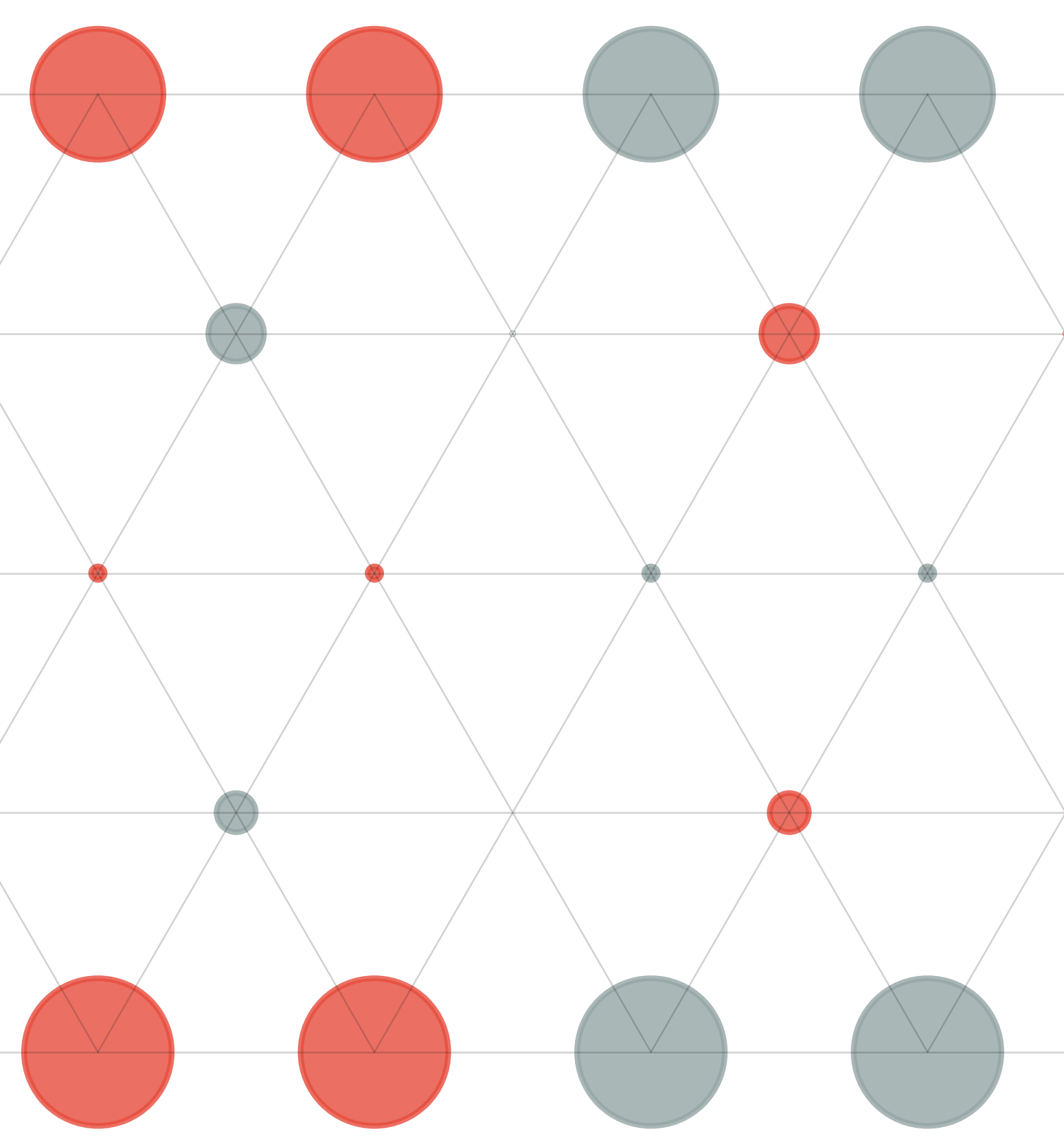} \hspace{2mm} \includegraphics[width = 3cm]{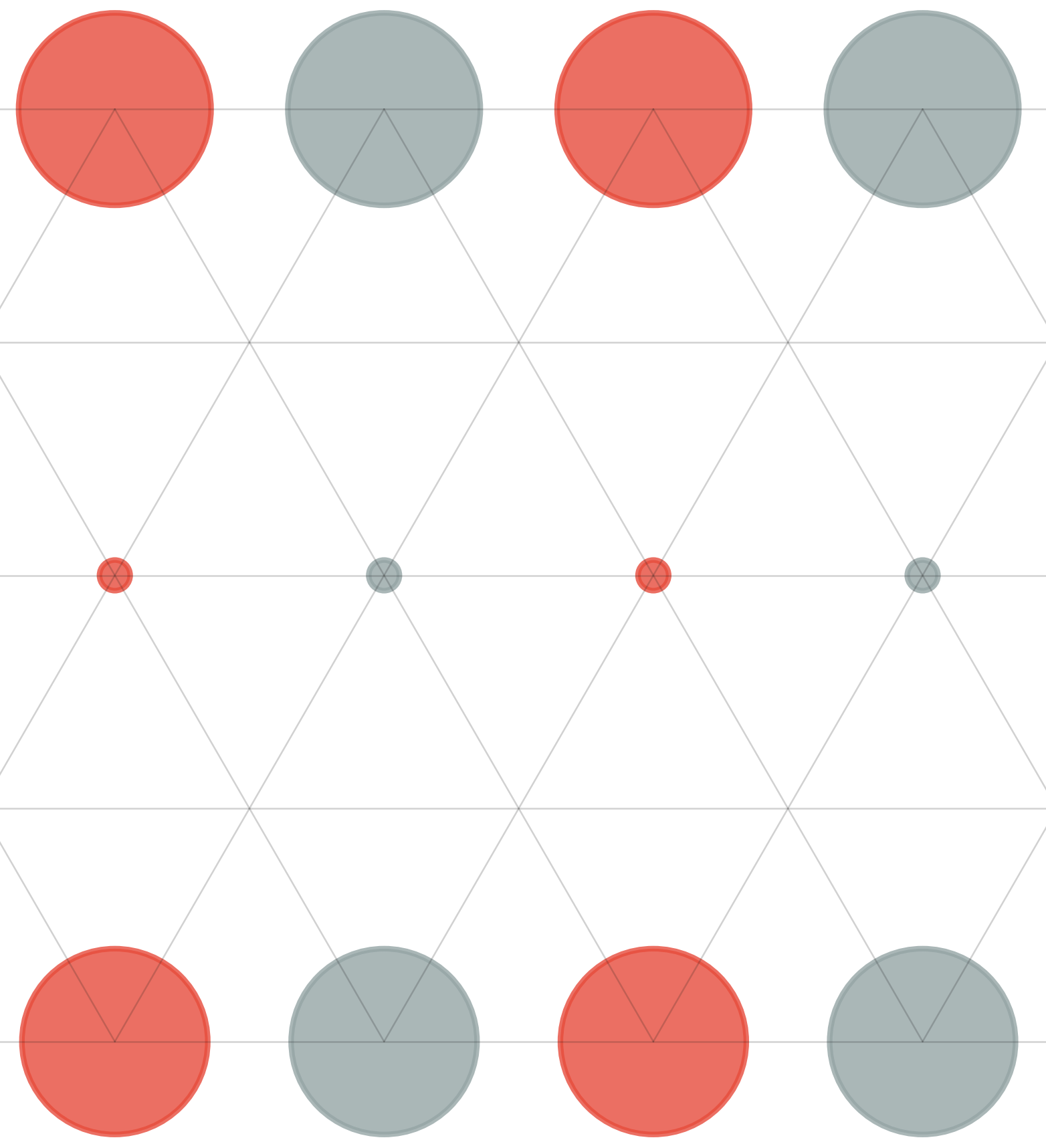}

\caption{(Top) Variation of the mean field staggered magnetizations
at the M-edge (orange) and the X-edge (blue) for the AF1 phase for
different TMD nanoribbons, all of them with a width of 5 $M$ atoms and with $U = 2.94$ eV. (Bottom) Example of the gapped AF2 (left) and AF1 (right) phases obtained with mean field theory at $\nu_\text{edge} = 0.5$ for a $\text{WSe}_2$ nanoribbon with a width of 5 $M$ atoms and considering $U = 2.94$ eV. We also recover the edge ferromagnetic phases for $\nu_\text{edge} \neq 0.5, 0.75$ shown in Fig. \ref{fig:mf_phases} for all TMDs in the family, but we omit them here for the sake of brevity.
\label{fig:stMagComp}}
\end{figure}

We obtain the edge antiferromagnetic phases AF2 and AF1, respectively at $\nu_\text{edge} = 0.5, 0.75$ for all the TMDs we considered. For other edge fillings, we find edge ferromagnetism. Even though there are slight differences in the specific values of the order parameters on each edge between the TMDs, our general qualitative conclusions do not change.  We illustrate this for the AF1 phase in the top panel of Fig.~\ref{fig:stMagComp}. Notice that for $\text{MoSe}_2$ and $\text{MoTe}_2$ the two edges have the same value of the staggered magnetization. This is because these two TMDs have the narrowest edge bandwidths. Thus, for this specific Hubbard-$U$, the magnetizations are likely already saturated, i.e. they have already reached their maximum value. On the bottom panel of Fig.~\ref{fig:stMagComp}, we show an example of the aforementioned edge antiferromagnetic phases for $\text{WSe}_2$.

\section{Determinant Quantum Monte Carlo method\label{sec:B}}

This appendix contains a brief description of our implementation of the 
DetQMC method that was used throughout this work and is publicly available at \footnote{DetQMC implementation, \url{https://github.com/fmonteir/tmd-nanoribbon-detQMC}, [Accessed: 20-March-2022]}. Extensive details and bench-marking of the implementation are also available in Ref.\cite{mastersthesis}. In order to enable the reader to reproduce our results, we focus on the aspects that are most relevant to this work. In particular, we provide details on how to write the spin correlation operator in
terms of the Green's functions, which are the main object of DetQMC simulations.
In theory, expectations of quantum observables can be computed directly from the partition function. Unfortunately, it is not possible to obtain an explicit closed form expression for the latter for the interacting systems in this work. Thus DetQMC makes use of Monte Carlo sampling to compute expectations of the spin correlation operator.

In the path integral formulation, with discretized
imaginary time, the partition function contains a product of exponential functions of a sum of non-commuting operators. This product can be approximated by using the Trotter breakup. Dividing the imaginary time interval $[0,\beta]$ into $L$ equal sub-intervals of smaller width $\Delta\tau=\beta/L$, and using the inverse of the Baker--Campbell--Hausdorff formula, whilst keeping only the first order term in $\Delta\tau$, we obtain

\begin{equation}
Z=\text{Tr}\bigg[\prod_{l=0}^{L-1}e^{-\Delta\tau\mathcal{H}_\text{TB}}e^{-\Delta\tau\mathcal{H}_{U}^{l}}\bigg]+\mathcal{O}(\Delta\tau^{2}),\label{eq:trotter}
\end{equation}
where $\mathcal{{H}}_\text{TB}$ is the three-band tight-binding Hamiltonian and $\mathcal{{H}}_{U}^{l}$ is the intraorbital Hubbard term defined in
the Hilbert space of the $l$-th imaginary time slice. The parameter $\Delta\tau^{-1}$ can be regarded as a high energy cutoff, and it must be larger than all other energy scales in the problem for the approximation to be valid.

Let us define the so called Hubbard Stratonovich (HS) binary field $\bm{h}$ as a $(L\times N)$-dimensional, spin-$1/2$ field comprised of binary variables. The interaction term is eliminated by use of the discrete HS transformation for $U>0$ \cite{bai_numerical_2009,hirsch_discrete_1983}
. Since $[n_{i,\alpha,\sigma},n_{j,\beta,\sigma'}]=0\,\,\forall i,j,\alpha,\beta,\sigma,\sigma'$,
we have 
\begin{equation}
e^{-\Delta\tau\mathcal{H}_{U}}=\prod_{i,\alpha}e^{-U\Delta\tau(n_{i,\alpha,\uparrow}-1/2)(n_{i,\alpha,\downarrow}-1/2)}.\label{eq:Hint}
\end{equation}

We will recast Eq.(\ref{eq:Hint}) in terms of the local electronic spin $n_{i,\alpha,\uparrow}-n_{i,\alpha,\downarrow}$, yielding a non-interacting quadratic term. Let $C=\frac{1}{2}e^{-\frac{U\Delta\tau}{4}}$ and $\nu=\text{arcosh}(e^{\frac{U\Delta\tau}{2}})$. Introducing the binary variables $\widetilde{h}_{i,\alpha}=\pm1$, the discrete HST

\begin{equation}
e^{-U\Delta\tau(n_{i,\alpha,\uparrow}-1/2)(n_{i,\alpha,\downarrow}-1/2)}=C\sum_{\widetilde{h}_{i,\alpha}}e^{\nu\widetilde{h}_{i,\alpha}(n_{i,\alpha,\uparrow}-n_{i,\alpha,\downarrow})},\label{eq:discreteHS}
\end{equation}
allows one to write the exponential of the Hubbard term as a trace
over the field at imaginary time slice $l$ \cite{bai_numerical_2009}. In principle, a more complicated transformation could allow one to simulate the model with interorbital, Hund and pair-hopping terms. However, it would require three spin-$1$ fields \cite{huang_sign-free_2022}, thereby significantly increasing the computational cost, which is already quite high due to the sign problem. Moreover, such a transformation would likely lead to a more severe sign problem \cite{held_microscopic_1998}, increasing the computational cost even more or impeding simulations altogether. For the sake of simplicity and to avoid excessive computational cost, we have only included the intraorbital term in our simulations.

Let us define $\mathcal{H}_{U,\sigma}=\sum_{i,\alpha}\nu\widetilde{h}_{i,\alpha}n_{i,\alpha,\sigma}=\sigma\nu\bm{c}_{\sigma}^{\dagger}\bm{U}(\widetilde{\bm{h}})\bm{c}_{\sigma}$, with $\bm{U}(\widetilde{\bm{h}})\equiv\text{diag}(\widetilde{h}_{i,\alpha})$.
Now, define HS fields for each imaginary time slice $\widetilde{\bm{h}_{l}}$,
which in turn specifies $\bm{U}_{l}$ and $\mathcal{H}_{U,\sigma}^{l}$.
Including the trace over the field and exchanging it with the fermionic
trace in Eq.(\ref{eq:trotter}), we obtain

\begin{equation}
Z=C^{NL}\text{Tr}_{\bm{h}}\text{Tr}\bigg[\prod_{l=0}^{L-1}\underbrace{e^{-\Delta\tau\mathcal{H}_{\text{TB}, \uparrow}}e^{\mathcal{H}_{U, \uparrow}^{l}}}_{B_{l,\uparrow}(\widetilde{\bm{h}_{l})}}\underbrace{e^{-\Delta\tau\mathcal{H}_{\text{TB} \downarrow}}e^{\mathcal{H}_{U,\downarrow}^{l}}}_{B_{l,\downarrow}(\widetilde{\bm{h}_{l})}}\bigg],
\end{equation}
where all operators are now quadratic in the fermion operators.
For the latter, the trace over the electronic
degrees of freedom may be taken explicitly \cite{hanke_electronic_1993},
turning the many-fermion problem into a single-particle problem:

\begin{equation}
Z=C^{NL}\text{Tr}_{\bm{h}}\bigg[\text{\ensuremath{\prod_{\sigma}}det}[\bm{I}+\prod_{l=L-1}^{0}\bm{B}_{l,\sigma}(\widetilde{\bm{h}}_{l})]\bigg].
\end{equation}

To multiply the chains of $\bm{B}$-matrices in a numerically stable manner, we use QR decompositions with partial pivoting \cite{hanke_electronic_1993,bai_numerical_2009,bai_stable_2011}.
The determinant can be calculated in $\mathcal{O}(LN^{3})$ flops
for a matrix whose size is polynomial in the number of sites $N$,
leading to a naive $\mathcal{O}(L^{2}N^{4})$ algorithm. To sample
configurations of $\bm{h}$, we use single spin-flip dynamics. The
acceptance/rejection scheme of the Metropolis algorithm is implemented
using a rank-one update of the matrices $\bm{I}+\prod_{l=L-1}^{0}\bm{B}_{l,\sigma}(\widetilde{\bm{h}}_{l})$
\cite{bai_numerical_2009}, which reduces the complexity of the algorithm
to order $\mathcal{O}(LN^{3})$. Using Wick's theorem,
we may write any observable in terms of the matrix elements of the
Green's function for a fixed configuration of the HS field, which
in turn is given by $\bm{G}^{\sigma}(\bm{h})=[\bm{I}+\prod_{l=L-1}^{0}\bm{B}_{l,\sigma}(\widetilde{\bm{h}}_{l})]^{-1}\,\,$
\cite{hanke_electronic_1993,bai_numerical_2009}.
We use the Green's function --- the fundamental object of DetQMC --- to sample configurations of the field $\bm{h}$ and to measure spin correlations. We do so by averaging 
the spin correlation operator over uncorrelated configurations of the HS field $\bm h$ to obtain an estimator of $\langle S_{i,\alpha}^{z}S_{j,\beta}^{z}\rangle $, the spin correlation between site/orbital pairs  $i, \alpha$ and $j, \beta$. For each configuration, we measure the observable $\langle S_{i,\alpha}^{z}S_{j,\beta}^{z}\rangle _{\mathbf{\bm{h}}}$,
defined in terms of $\bm{G}^{\sigma}(\bm{h})$ as

\begin{widetext}
\begin{equation}
\left\langle S_{i,\alpha}^{z}S_{j,\beta}^{z}\right\rangle_{\bm h} =\begin{cases}\sum_{\sigma}\Big(G_{(i\alpha)(i\alpha)}^{\sigma}(\bm{h}) G_{(j\beta)(j\beta)}^{\sigma}(\bm{h}) -G_{(i\alpha)(i\alpha)}^{\sigma} (\bm{h}) G_{(j\beta)(j\beta)}^{-\sigma} (\bm{h}) \Big) & (i\alpha)\neq(j\beta)\\\sum_{\sigma}G_{(i\alpha)(i\alpha)}^{\sigma}(\bm{h})-2G_{(i\alpha)(i\alpha)}^{\uparrow}(\bm{h})G_{(i\alpha)(i\alpha)}^{\downarrow}(\bm{h}) & (i\alpha)=(j\beta)\end{cases}.
\end{equation}
\end{widetext}

A final remark must be made about computational effort. These simulations are plagued by the sign problem --- which exponentially increases the variance of our estimators --- deeming them very computationally intensive. In order to give the reader a concrete idea of just how intensive these simulations are, we compared two of the points shown in Fig.~\ref{fig:qmc_sign}: we fixed $N_x = 20$ and compared the data points shown for $U = 2.76, 2.94\,eV$. Since the average sign for $U = 2.94\,eV$ ($\langle \text{sign} \rangle = 0.2963 \pm 0.0007$) is lower than for $U = 2.76\,eV$ ($\langle \text{sign} \rangle = 0.470 \pm 0.002$), we expected to need more CPU hours in the case of the former in order to obtain similar accuracy to the case of the latter. This expectation was confirmed: 2930 CPU hours were required in order to obtain an error of $ \Delta n = 0.005$ in the electron density $\langle n \rangle$ for $U = 2.76\,eV$, whilst 26093 CPU hours (almost 9 times more) were required in order to obtain an error of $\Delta n = 0.003$ for $U = 2.94\,eV$. The data points $N_x = 20,\,U = 2.76,\, 2.94\,eV$ were chosen to illustrate the computational cost because they were some of the most statistically demanding parts of this study. 

\bibliographystyle{apsrev4-2}
\bibliography{references}

\end{document}